# Probabilistic "Copies" in Generative AI Models

Mark A. Lemley* & A. Feder Cooper**

ABSTRACT

*Recent work shows that it is possible to extract verbatim or near-verbatim text of some copyrighted works from some large language models (LLMs or models). That is evidence that the model weights encode the works in some form—that the model has "memorized" those works from its training data. But LLMs don't store information in the same format as familiar databases. Rather, their weights store statistical relationships that have been learned from the training data, and those relationships inform a generation process that in practice is often probabilistic rather than deterministic. In the case of memorization, those relationships are strong enough that, in many circumstances, the model could generate a copyrighted work from its training data.*

*Copyright law has not previously had to decide whether storing information that might or might not produce output similar to a copyrighted work is itself a copy of the work. The answer to this question is important, because it may determine the legality of many LLMs. The statute and caselaw are largely unhelpful. We argue that copyright law will likely take a functional approach to the question, finding that LLMs contain a copy of a particular work only if it is straightforward to extract that work in outputs. That result is unsatisfying as a policy matter, and we suggest potential changes to the law, but it is the most likely outcome under current law.*

---

* William H. Neukom Professor, Stanford Law School.

** Assistant Professor of Computer Science, Yale University. The authors contributed equally to this work. Thanks to Paul Goldstein, James Grimmelmann, Rose Hagan, Jessica Litman, Tyler Ochoa, Tony Reese, Todd Smith, Xiyin Tang, Rebecca Tushnet, participants at the Copyright Act at 50 conference at Stanford Law School, and participants at the 29th Annual BTLJ-BCLT Spring Symposium: Origins, Evolution, and Possible Futures of the 1976 Copyright Act for conversations on this topic and to Austin Taylor for outstanding research assistance. We note that one of us (Lemley) has in the past represented the defendants in some of the cases discussed here, including *Andersen v. Stability AI*, *Kadrey v. Meta*, and *Bartz v. Anthropic*. As of July 2026 he is not working for any clients. Our work is our own and does not reflect the views of any company or any confidential information obtained from any company.

TABLE OF CONTENTS

## I.
## INTRODUCTION

More than one hundred pending lawsuits assert that generative AI infringes copyright law.[1] Most of those suits challenge the legality of training AI on copyrighted works.[2] An increasing number allege infringement because the output of AIs is the same as or similar to a copyrighted work,[3] for instance, a near-exact reproduction of specific song lyrics.[4]

1. David Atkinson, *AI Litigation Tracker—AtkinsonAILaw.com*, GOOGLE SHEETS, https://docs.google.com/spreadsheets/d/1BgoFqE0C6148M2evzZU9A-WJN0oys7Pm [https://perma.cc/AWG7-G88F] (last visited Aug. 29, 2026); *see also ChatGPT is Eating the World*, CHAT GPT IS EATING THE WORLD, https://chatgptiseatingtheworld.com/ [https://perma.cc/LDY3-TP7L] (last visited Aug. 29, 2026) (providing live count of total U.S. cases related to AI litigation).

2. *See, e.g.*, Bartz v. Anthropic PBC, 787 F. Supp. 3d 1007, 1034 (N.D. Cal. 2025) (holding that copying of works to train LLMs is fair use); Kadrey v. Meta Platforms, Inc., 788 F. Supp. 3d 1026, 1057 (N.D. Cal. 2025) (holding that copying of works to train LLMs is fair use, but emphasizing that insufficient evidence was presented by the plaintiffs on the effect of use upon the potential market for the copyrighted works); Thomson Reuters Enter. Ctr. GmbH v. Ross Intel. Inc., 765 F. Supp. 3d 382, 400–01 (D. Del. 2025) (holding that copying of works to train a legal research platform on Reuters' headnotes was not protected by fair use primarily because Ross was developing a market substitute).

3. *See, e.g.*, N.Y. Times Co. v. Microsoft Co., 777 F. Supp. 3d 283, 328 (S.D.N.Y. 2025) (denying motion to dismiss plaintiffs' claim that defendants' LLMs outputs were substantially similar to plaintiffs' works); *In re* OpenAI, Inc., Copyright Infringement Litig., 800 F. Supp. 3d 602, 612–13 (S.D.N.Y 2025) (declining to dismiss plaintiffs' claims that some model outputs were substantially similar to plaintiffs' works); Tremblay v. OpenAI, Inc., 716 F. Supp. 3d 772, 778, 780 (N.D. Cal. 2024) (dismissing plaintiffs' claim that every output of the model trained on their copyrighted materials is an infringing derivative work).

4. Landgericht München I [LG München I] [Munich Regional Court I], Nov. 11, 2025, Gewerblicher Rechtsschutz und Urheberrecht [GRUR] 1917 (2025) (Ger.) (holding that reproduction of song lyrics in model outputs was infringing).

Until November 2025, courts and litigants hadn't paid much attention to what happens in between training the model and generating outputs—the copyright status of the model itself. A few early U.S. suits claimed the models were derivative works simply because they were trained on copyrighted works. Courts generally rejected those claims.[5] But two recent decisions—one in Germany and one in the U.K.—have brought the question of whether the model itself incorporates a copy of the works it was trained on to the fore. One decision found that ChatGPT incorporated a copy of copyrighted song lyrics, which appeared in outputs.[6] The other, involving the image-generation model Stable Diffusion, found the opposite.[7] U.S. courts have not yet confronted the issue of whether the model is a copy.[8]

The question is likely to grow in importance and touches on a line of technical research that the scientific literature refers to as "memorization." We address the relevant details of this term in Part I, but the intuition is as follows: during training, specific training data leave a sufficiently strong imprint on the model's weights such that it can be possible to reproduce—i.e., extract—those training data in the model's outputs. In other words, "extraction" is an observable consequence of memorization: if a piece of training data is memorized, it may be possible to extract it from the model.[9]

5. *See* Andersen v. Stability AI Ltd., 700 F. Supp. 3d 853, 868–69 (N.D. Cal. 2023) (dismissing plaintiffs' claim that the model itself is an infringing derivative work); *Kadrey*, 788 F. Supp. 3d at 1056–57 (dismissing plaintiffs' claim that the model itself is an infringing derivative work).

6. *See* GRUR 1917 (2025) (Ger.); Jörn Poltz & Friederike Heine, *OpenAI Used Song Lyrics in Violation of Copyright Laws, German Court Says*, REUTERS (Nov. 11, 2025), https://www.reuters.com/world/german-court-sides-with-plaintiff-copyright-case-against-openai-2025-11-11 [https://perma.cc/6PQ2-ZM97] (summarizing some of the plaintiff material from *GEMA v. OpenAI*).

7. *See generally* Getty Images (US) Inc v. Stability AI Limited [2025] EWHC (Ch) 2863 (Eng.) (holding that Stability AI's model was not itself an infringing copy of Getty's images).

8. Some courts have rejected claims that the model is by definition a derivative work of anything it is trained on. *See Andersen*, 700 F. Supp. 3d at 861. There is a natural relationship between the training of a model and any content that is memorized in the model. But the training cases do not present the question of whether a copy has been made in the model; rather, they concern the intermediate copy which defendants concede is contained in the training data, and defend as fair use.

9. *See* A. Feder Cooper & James Grimmelmann, *The Files are in the Computer: On Copyright, Memorization, and Generative AI*, 100 CHI.-KENT L. REV. 141, 142 (2025); A. Feder Cooper et al., Report of the 1st Workshop on Generative AI and Law 30 (Dec. 3, 2023) (unpublished manuscript) [hereinafter GenLaw Report], https://arxiv.org/pdf/2311.06477 [https://perma.cc/S3M6-YUF2]. Memorization is a technical term in the scientific literature on machine learning, not one that we have picked to describe the phenomenon in question. In many respects, the choice of terminology is an unfortunate one, as it sounds like a metaphor for how AI models behave—that they "memorize" information in a way like a human might when studying for an exam. But this term is not a metaphor, and it isn't meant to invoke this meaning: it is shorthand for the specific technical phenomenon that we provide an intuition for here, and for which we provide more background details below. *See infra* Section II.B.

Researchers in machine learning have studied both memorization[10] and extraction[11] for years, long before generative AI attracted attention from copyright law. But the training of commercial AI systems on copyrighted works has made memorization a copyright question, not just a machine learning one. Academic research by ourselves and others has shown that it's possible to extract at least some copyrighted works verbatim from AI models.[12] More recent work shows that, even if a copyrighted work can't be extracted verbatim from an AI model, the model can sometimes reproduce near-verbatim copies in its outputs.[13] These copies are sufficiently similar that they would infringe copyright, which does not require identical duplication.[14]

10. *See, e.g.*, Vitaly Feldman, Does Learning Require Memorization? A Short Tale about a Long Tail (Jan. 10, 2021) (unpublished manuscript), https://arxiv.org/pdf/1906.05271 [https://perma.cc/Q4VH-JSW4]; Reza Shokri, Marco Stronati, Congzheng Song & Vitaly Shmatikov, Membership Inference Attacks against Machine Learning Models (Mar. 31, 2017) (unpublished manuscript), https://arxiv.org/pdf/1610.05820 [https://perma.cc/F5JR-U36D].

11. *See, e.g.*, Nicholas Carlini, Florian Tramèr, Eric Wallace, Matthew Jagielski, Ariel Herbert-Voss, Katherine Lee, Adam Roberts, Tom Brown, Dawn Song, Úlfar Erlingsson, Alina Oprea & Colin Raffel, *Extracting Training Data from Large Language Models* [hereinafter *Extracting Training Data from LLMs*], *in* 30TH USENIX SECURITY SYMPOSIUM 2633, 2633 (USENIX Ass'n 2021); Katherine Lee, Daphne Ippolito, Andrew Nystrom, Chiyuan Zhang, Douglas Eck, Chris Callison-Burch & Nicholas Carlini, *Deduplicating Training Data Makes Language Models Better* [hereinafter *Deduplicating*], *in* 1 PROCEEDINGS OF THE 60TH ANNUAL MEETING OF THE ASSOCIATION FOR COMPUTATIONAL LINGUISTICS 3 (Smaranda Muresan, Preslav Nakov & Aline Villavicencio eds., 2022).

12. *See* Milad Nasr, Nicholas Carlini, Jonathan Hayase, Matthew Jagielski, A. Feder Cooper, Daphne Ippolito, Christopher A. Choquette-Choo, Eric Wallace, Florian Tramèr & Katherine Lee, Scalable Extraction of Training Data from (Production) Language Models 9 (Nov. 28, 2023) (unpublished manuscript), https://arxiv.org/pdf/2311.17035 [https://perma.cc/4P6L-VDC8]; A. Feder Cooper, Mark A. Lemley, Allison Casasola, Ahmed Ahmed, Aaron Gokaslan, Amy B. Cyphert, Christopher De Sa, Daniel E. Ho & Percy Liang, *Extracting Memorized Pieces of (Copyrighted) Books from Open-Weight Language Models*, 3 CONF. LANGUAGE MODELING (forthcoming Oct. 2026) (manuscript at 14) (on file with authors) [hereinafter *Extracting Books (Open-Weight)*]; Antonia Karamolegkou, Jiaang Li, Li Zhou & Anders Søgaard, *Copyright Violations and Large Language Models*, *in* PROCEEDINGS OF THE 2023 CONFERENCE ON EMPIRICAL METHODS IN NATURAL LANGUAGE PROCESSING 7403, 7406 (Houda Bouamor, Juan Pino & Kalika Bali eds., 2023).

13. *See* Ahmed Ahmed, A. Feder Cooper, Sanmi Koyejo & Percy Liang, Extracting Books from Production Language Models 4 (Jan. 6, 2026) (unpublished manuscript), https://arxiv.org/pdf/2601.02671 [https://perma.cc/U987-KELD] [hereinafter Extracting Books (Production LLMs)]; Guangwei Zhang, Jianing Zhu, Cheng Qian, Neil Gong, Rada Mihalcea, Zhaozhuo Xu, Jingrui He, Jiaqi Ma, Yun Huang, Chaowei Xiao, Bo Li, Ahmed Abbasi, Dongwon Lee, Heng Ji & Denghui Zhang, Copyright Detective: A Forensic System to Evidence LLMs Flickering Copyright Leakage Risks 7 (Feb. 11, 2026) (unpublished manuscript), https://arxiv.org/pdf/2602.05252 [https://perma.cc/G3QG-YFWW]; A. Feder Cooper, Mark A. Lemley, Christopher De Sa, Lea Duesterwald, Allison Casasola, Jamie Hayes, Katherine Lee, Daniel E. Ho & Percy Liang, *Estimating Near-Verbatim Extraction Risk in Language Models with Decoding-Constrained Beam Search*, 3 CONF. LANGUAGE MODELING (forthcoming 2026) (manuscript at 1) (on file with authors) [hereinafter *Near-Verbatim*].

14. "[A]n infringement is not confined to literal and exact repetition or reproduction; it includes also the various modes in which the matter of any work may be adopted, imitated, transferred, or reproduced, with more or less colorable alterations to disguise the piracy." Universal Pictures Co. v. Harold Lloyd Corp., 162 F.2d 354, 360 (9th Cir. 1947). The test for copyright infringement is substantial similarity, though how that similarity is tested can differ by circuit. The Ninth Circuit, for example, uses a two-part test: an extrinsic test based on objective similarity, and an intrinsic test based on the response

These findings have implications for AI outputs, but also potentially for the models themselves. As is well-established in the AI scientific community, extraction is evidence that the memorized training data is encoded in some form in the model's weights.[15] Writing about this for a legal audience, Cooper and Grimmelmann note that the fact that a model can generate long verbatim snippets of a book (e.g., text from *Harry Potter*) means that at least those portions of the book are encoded in the model itself.[16] And the Copyright Office has declared, without case support, that because model weights are fixed in a tangible medium, anything contained in them is also a fixed copy.[17]

But the *extent* and the *form* in which that encoding happens are likely to matter a great deal under existing copyright law. If an AI model contains a "copy" of a copyrighted work as copyright law defines that term, and it merely replicates that copy in its outputs on request, the creation of the model itself is an act of direct infringement; so is the making of a new copy of that model—for example, downloading the weights of an openly released model, like one from Meta's Llama 3 series, to a corporate server. That is true even if no one ever asks for—and the model never generates—infringing output.[18] By contrast, if the model contains only information that could be used to produce a copy and generates the output only in response to particular prompts—assembling the output on the fly from sources that do not include a legally cognizable copy of of a lay observer. *See* Sid & Marty Krofft Television Prods., Inc. v. McDonald's Corp., 562 F.2d 1157, 1164 (9th Cir. 1977), *overruled on other grounds by*, Skidmore as Tr. for Randy Craig Wolfe Tr. v. Led Zeppelin, 952 F.3d 1051 (9th Cir. 2020). The Second Circuit, similarly, asks "whether defendant took from plaintiff's works so much of what is pleasing to the ears of lay listeners." Arnstein v. Porter, 154 F.2d 464, 473 (2d Cir. 1946); *see* Mark A. Lemley, *Our Bizarre System for Proving Copyright Infringement*, 57 J. COPYRIGHT SOC'Y 719 (2010) (discussing the various tests).

15. For discussions of extraction as evidence of memorization, see, for example, *Extracting Training Data from LLMs*, *supra* note 11; Nicholas Carlini, Jamie Hayes, Milad Nasr, Matthew Jagielski, Vikash Sehwag, Florian Tramèr, Borja Balle, Daphne Ippolito & Eric Wallace, *Extracting Training Data from Diffusion Models* [hereinafter *Extracting Training Data from Diffusion Models*], *in* 32ND USENIX SECURITY SYMPOSIUM 5253 (USENIX Ass'n 2023); Nicholas Carlini, *What My Privacy Papers (Don't) Have to Say About Copyright and Generative AI* (Mar. 11, 2025) [hereinafter *Carlini Blog*], https://nicholas.carlini.com/writing/2025/privacy-copyright-and-generative-models.html [https://perma.cc/R34S-W8SM]; Avi Schwarzschild, Zhili Feng, Pratyush Maini, Zachary C. Lipton & J. Zico Kolter, Rethinking LLM Memorization Through the Lens of Adversarial Compression (Nov. 11, 2024) (unpublished manuscript), https://arxiv.org/pdf/2404.15146 [https://perma.cc/UBX4-9BFD]; Nasr et al., *supra* note 12; and Giorgio Franceschelli, Claudia Cevenini & Mirco Musolesi, Training Foundation Models as Data Compression: On Information, Model Weights and Copyright Law (Mar. 12, 2025) (unpublished manuscript), https://arxiv.org/pdf/2407.13493 [https://perma.cc/QB4M-HTFV].

16. *See* Cooper & Grimmelmann, *supra* note 9, at 168.

17. U.S. COPYRIGHT OFF., COPYRIGHT AND ARTIFICIAL INTELLIGENCE: PART 3: GENERATIVE AI TRAINING 28–29 (2025), https://www.copyright.gov/ai/Copyright-and-Artificial-Intelligence-Part-3-Generative-AI-Training-Report-Pre-Publication-Version.pdf [https://perma.cc/3FZG-33HJ]; *see also* Daniel Gervais, Haralambos Marmanis, Noam Shemtov & Catherine Zaller Rowland, *The Heart of the Matter: Copyright, AI Training, and LLMs*, 71 J. COPYRIGHT SOC'Y 482, 494–98 (2024); Joshua Yuvaraj, *Pseudo-Expressions: An Expanded Conception of the Work in Copyright Law* (Working Paper 2025) (both arguing that model weights can be copies).

18. The copy in the model in this scenario may still qualify as fair use; more on that below.

the work itself—the output may be an infringing copy but the model itself likely wouldn't be.

Determining which of these descriptions fits AI models turns out to be surprisingly complicated. The difficulty lies both in the technical details of how models encode information and generate outputs, and in copyright law's existing treatment of copies and fixation.

The rest of this Article proceeds in two parts. Part I provides a minimal account of how language models encode information, how "memorization" reflects a particular degree of encoding of training data in the model's weights, and why ordinary metaphors like storage, memory, and compression both illuminate and mislead. The point isn't to settle the legal questions by analogy, but to show why the relevant copyright questions can't be resolved at the level of abstraction those analogies invite. Part II then turns to copyright law, examining whether models that can (deterministically or probabilistically) generate copyrighted works from their training data should be treated as fixed copies, and what follows if they are.

## II.
## THE TECHNICAL COMPLICATIONS OF MEMORIZATION AND EXTRACTION

What is often called "generative AI" refers to a set of technological systems that resist simple description. It's therefore natural to reach for metaphors and analogies—to use the familiar to try to make sense of things that are complex and unfamiliar. This is an especially common pattern in law: "Analogy and metaphor are central to legal rhetoric; they provide a rational framework for thinking through the relevant similarities and differences between cases."[19] But "metaphors . . . can also simplify and distort,"[20] and so legal practitioners (as well as others) "should beware of metaphors that draw on more familiar technologies, which provide inadequate answers to the genuinely hard problems before them."[21]

This warning is critical to keep front of mind when thinking about what the machine learning literature refers to as "memorization," as well as related terms such as "compression" and "training." These terms all invoke familiar, everyday meanings, but in machine learning they also refer to precise technical concepts.[22]

19. GenLaw Report, *supra* note 9, at 4; *see also* Daniel J. Solove, *Privacy and Power: Computer Databases and Metaphors for Information Privacy*, 53 STAN. L. REV. 1393, 1397–99 (2001) (arguing that the database problem cannot be adequately understood through the use of the "Big Brother metaphor").

20. GenLaw Report, *supra* note 9, at 4; *see also id.* at Appendix B; Katherine Lee, A. Feder Cooper & James Grimmelmann, *Talkin' 'Bout AI Generation: Copyright and the Generative-AI Supply Chain*, 72 J. COPYRIGHT SOC'Y 251, Part III (2025) [hereinafter *Talkin'*]. For a longer and more recent treatment of related issues, see Michiel A. Smit, *Metaphors We Judge (AI) By: A Rhetorical Analysis of Artificial Copyright Disputes*, 21 J. INTELL. PROP. L. & PRAC. 251 (2026) (Eng.).

21. *See Talkin'*, *supra* note 20, at 255–59.

22. *See* GenLaw Report, *supra* note 9, at 4–5.

A central challenge in work at the intersection of generative AI and law is that the scientific literature is rife with terms that do this metaphorical-technical double duty: shorthand labels that gesture toward, but also obscure, the underlying technical mechanisms they are meant to describe.[23] If one, ironically, leans too literally into the metaphorical side, it can be tempting to reject these terms as "fallacies" or failures of abstraction.[24] But we can't understand the accuracy of the metaphor without the technical substance those terms are actually referring to.

At the risk of making our own imperfect analogy, law faces similar terminological challenges. Many core legal concepts are easily misunderstood by non-experts because they carry ordinary meanings that differ from their technical legal definitions. In copyright, "fair use," "transformative," and even "copy" are three such concepts that invite confusion when invoked on non-technical terms. But the solution in law isn't to throw out these terms—to call them "fallacies" because they are frequently misunderstood by others—but to engage them thoughtfully on their technical merits to make sound, reasoned arguments.

So in that spirit, the remainder of this Section develops a minimal technical account of language models and "memorization"—when the imprint left on the model's weights by a specific work in the training data is sufficient to encode part or all of a copyrighted work in the model itself. We focus on language models because they have been the subject of our own empirical work in machine learning, from which we draw clear examples of memorization.[25] We don't attempt to provide a comprehensive treatment of memorization, and instead defer to others.[26] Rather, we aim to clarify the specific mechanisms necessary to understand the relationship between memorization—which is a property of a trained model itself—and when a model produces copies of its training data as outputs. Further, it's important to remember that memorization is a complex phenomenon, for which the scientific literature continues to evolve. One of the key takeaways from the literature is that it isn't possible to make broad, one-size-fits-all statements about memorization and AI. This is in part why popular metaphorical understandings of memorization fall short, and so we explain why the relevant questions for copyright can't be resolved at the level of abstraction that metaphors and analogies encourage. This provides the foundation for our copyright analysis in Part II.

---

23. Smit, *supra* note 20, at 251–52.

24. *See infra* Section II.C.

25. While we focus on text and LLMs, memorization also arises in other forms of generative AI, such as images and software code. *See generally Extracting Training Data from Diffusion Models*, *supra* note 15 (extracting images from Stable Diffusion); Aneesh Pappu, Billy Porter, Ilia Shumailov & Jaime Hayes, *Measuring Memorization in RLHF for Code Completion*, THIRTEENTH INT'L CONF. ON LEARNING REPRESENTATIONS, Feb. 25, 2025, art. 4775 (studying memorization in code completion). We particularly focus on *autoregressive* LLMs; while other LLMs, such as diffusion language models (DLMs) are growing in popularity, autoregressive LLMs currently remain the dominant paradigm.

26. *See generally* Cooper & Grimmelmann, *supra* note 9 (defining memorization for a legal audience); *see also* Lee et al., *supra* note 11, at 8 (discussing memorization in detail).

### *A. How Language Models Store Information and Generate Outputs*

Language models are made up of a large collection of numerical parameters, commonly referred to as weights. These weights are adjusted during training,[27] a process that takes as inputs both a model architecture (how those weights are structured and interact) and a training dataset. The result is a trained model: a fixed set of weights that encode information derived from the training data.[28]

This is how models store information[29] from their training data, and it's important to be clear about what that means. "Store" here does *not* literally mean that the model retains discrete copies of its training data in exactly the same format as a traditional database or lookup table would.[30] The information isn't localized or laid out in that way. But it does mean that certain information about the training data is encoded in the model's weights. That encoded information determines the probability distribution the model assigns to sequences of tokens that represent text. For example, after seeing the phrase "The dog barked at the" the model may assign a higher probability to the next token being "cat" compared to the next token being "chair" or "moon," reflecting what it has learned from the training data about the relationships between words.

27. The term "training" also performs the same metaphorical-technical double duty we've been talking about. In machine learning, it refers to an optimization process in which a model's weights are adjusted to minimize a loss function over a dataset, rather than to anything like human learning or instruction. Unlike "memorization" or "compression," however, "training" is largely understood in terms of its technical meaning in the legal literature. *See* GenLaw Report, *supra* note 9, at 4–5, 38.

28. *See Talkin'*, *supra* note 20, sections I.A and I.B for a treatment of background information on data and machine learning addressed to a legal audience.

29. Like "memorization" and "compression," the claim that language models "store" information is not merely metaphorical. In a technical sense, a trained model encodes information about its training data in its weights, which parameterize a probability distribution over token sequences. This concept can be formalized using tools from information theory: predictive models induce compression schemes in which better prediction corresponds to more efficient representation of the underlying data in the model's weights. *See generally* Grégoire Delétang, Anian Ruoss, Paul-Ambroise Duquenne, Elliot Catt, Tim Genewein, Christopher Mattern, Jordi Grau-Moya, Li Kevin Wenliang, Matthew Aitchison, Laurent Orseau, Marcus Hutter & Joel Veness, *Language Modeling Is Compression*, TWELFTH INT'L CONF. ON LEARNING REPRESENTATIONS, Mar. 14, 2024, art. 372. Of course, this notion of storage differs fundamentally from discrete, addressable storage in traditional databases or lookup tables, but it's nonetheless a precise and measurable property of the model and so it's correct to refer to this process as "storage." For more on why machine learning models are not at all like traditional databases, see generally A. Feder Cooper et al., *Machine Unlearning Doesn't Do What You Think: Lessons for Generative AI Policy and Research*, THIRTY-NINTH ANN. CONF. ON NEURAL INFO. PROCESSING SYS. POSITION PAPER TRACK, Jan. 16, 2026, art. 474 [hereinafter *Machine Unlearning*]. For a brief treatment of the relationship between compression and memorization, see generally Franceschelli et al., *supra* note 15.

30. The *format* is different from a traditional database or lookup table, but the *behavior* can be rigorously related to a stochastic lookup table. *See Machine Unlearning*, *supra* note 29, at 4–5.

Beyond this high-level description, we in fact know surprisingly little about how AI models actually store specific information. For a text file stored in a computer, one could point to a particular set of 1s and 0s and say "that's the data that corresponds to this particular file." That's generally not true for AI models. To be clear, the problem is *not* that the AI companies know this information and have concealed it from the world. It's that there remain big, open scientific questions on this topic: even the people who build and train models typically don't know precisely what pieces of information models retain or where exactly they store it.[31]

Due to this complexity, it's common to describe this whole process as learning "statistical correlations,"[32] "relationships,"[33] or "patterns"[34] in the training data. That description isn't wrong, but it's also only a very high-level gloss that oversimplifies the information represented by the model. For that matter, the same is true of saying that a model "learns" from its training data. That terminology is useful, but it shouldn't be taken too literally: a language model isn't a human brain, and its internal representation doesn't map exactly onto familiar concepts like human memory or understanding.

This is also the sense in which the term compression is sometimes used in the technical literature. Training a model can be understood as constructing a compact (i.e., compressed) representation of the data that supports accurate prediction.[35] This is almost surely what Emad Mostaque, former CEO of the Stability AI, had in mind when he said the following in 2022 about the Stable Diffusion image generation model:

> It's worth taking a step back and thinking about how crazy insane this is: we took a hundred terabytes of data—a hundred thousand thousand [sic] megabytes of images—2 billion of them—and we squished it down to a 2–4 gigabyte file. And that file can create everything that you've seen. That's insane, right? That's about as compressed as you can get.[36]

Compression is a technically accurate way to describe the results of training to a computer scientist; training a language model is compression in an information theory sense,[37] even though a model is not arranged or otherwise

31. Developing a better understanding of these phenomena is an active area of ongoing research in machine learning. *See generally Machine Unlearning*, *supra* note 29 and citations therein concerning mechanistic interpretability.

32. *See* Defendant Anthropic PBC's Opposition to Plaintiffs' Motion for Preliminary Injunction at 4, Concord Music Grp., Inc. v. Anthropic PBC, 738 F. Supp. 3d 973 (M.D. Tenn. 2024) (No. 3:23-cv-01092), Dkt. No. 67.

33. *See* Bartz v. Anthropic PBC, 787 F. Supp. 3d 1007, 1017 (N.D. Cal. 2025).

34. *See* Concord Music Grp., Inc. v. Anthropic PBC, 772 F. Supp. 3d 1131, 1134 (N.D. Cal. 2025).

35. *See* Delétang et al., *supra* note 29, at 1–3. More on this below, see *infra* Section II.C.

36. *See* First Amended Complaint at 29, Andersen v. Stability AI Ltd., 744 F. Supp. 3d 956 (N.D. Cal. 2024) (No. 23-cv-00201-WHO), Dkt. No. 129.

37. *See generally* Delétang et al., *supra* note 29; Franceschelli et al., *supra* note 15.

formatted like more familiar data formats, such as MP3 or PDF files.[38] But what information theory means by compression may not be what lay users have in mind when they hear the term.[39]

Once trained, a language model can be used to generate text. Given an input prompt, an autoregressive LLM predicts the next token in a sequence based on the tokens that precede it, then repeats this process iteratively to produce longer outputs.[40] Each prediction is probabilistic: based on the "patterns" the model has learned during training, it assigns different probabilities to possible next tokens, conditioned on the prior context.[41] We provide a high-level illustration of this process in Figure 1. At each generation step, a token is then selected from this distribution, which determines whether the model continues "Harry" with "Potter," "Truman," or something else.

*Figure 1: Illustrating autoregressive generation.*
*The input prompt is "Mr. and Mrs. Durs," for which the (autoregressive) LLM produces a distribution over the next tokens in its vocabulary (all possible tokens it can generate). At each generation step, the decoding procedure illustrated here selects the highest-probability token as the one to generate. In this example, the top-ranked token has enormous probability at each step. The output after several steps is verbatim text from Harry Potter and the Sorcerer's Stone. Figure produced by the authors, based on results for running this prompt on Llama 3.1 70B with greedy decoding.*

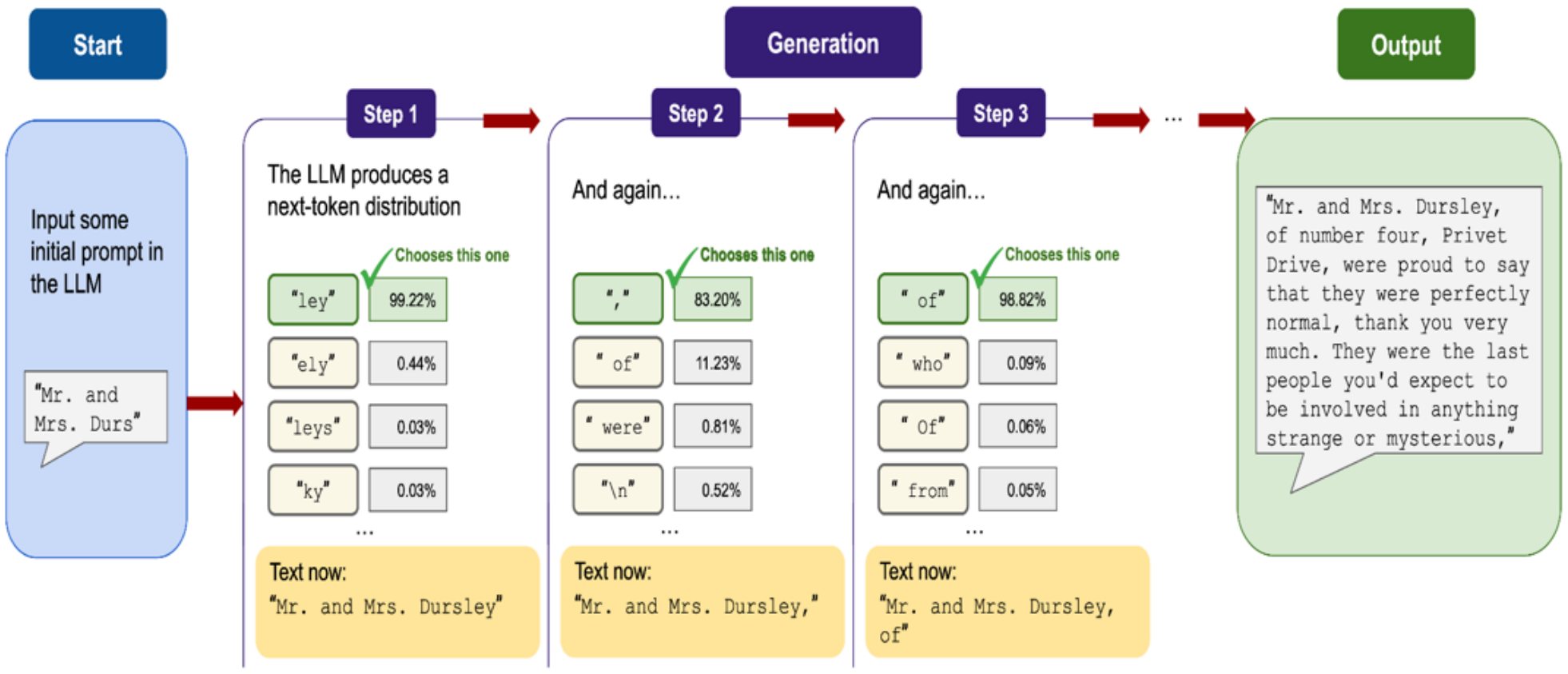

38. *See* Cooper & Grimmelmann, *supra* note 9, at 179–80.
39. Matthew Sag, *Copyright's Jagged Frontier*, DUKE L.J. (forthcoming 2026) (manuscript at 36), https://papers.ssrn.com/sol3/papers.cfm?abstract_id=6319379 (on file with the Berkeley Technology Law Journal) (noting the difference in how computer scientists use the term).
40. *See* Cooper & Grimmelmann, *supra* note 9, at 189. There are other types of language models, which produce outputs in different ways. Currently, autoregressive language models are the most common paradigm. *See supra* note 25.
41. *See* Cooper & Grimmelmann, *supra* note 9, at 189, 192–93. Hannibal Travis calls LLMs a lexicon of words, grammar, and concepts. *See* Hannibal Travis, *The Universal Lexicon: Generative AI Chatbots and Fair Use of Works in a Training Corpus*, 45 REV. LITIG. 61, 72–73 (2025).

This is where an important distinction arises between the model itself and the procedure used to generate outputs from it. The model defines a probability distribution over sequences. A given model is, in that sense, fixed once training is complete.[42] But generating text from that distribution requires a decoding procedure—a rule for selecting tokens from the model's predicted probabilities. In practice, decoding is often stochastic. Tokens are sampled from the probability distribution represented by the model, which means that the same prompt can yield different outputs on different runs. During generation, sometimes the model might produce "Potter" after "Harry," and sometimes it might produce "Truman"; how often it produces either depends on the learned distribution in the model weights.

But decoding need not be stochastic: there are also deterministic decoding procedures. For instance, greedy decoding always selects the highest-probability token from the next-token distribution at each generation step. Because the learned probability distribution is fixed, at each generation step for a given prompt, that top-ranked token is the same every time. There is no randomness involved in selecting it, and the model produces the same output every time for the given prompt.[43]

This distinction explains both the variability of model outputs and the limits of familiar analogies. Because the model itself defines a fixed probability distribution, the variability in outputs does *not* arise from changes in the model, but from how that distribution is used during decoding. The decoding procedure

42. It is fixed with respect to training being completed at the moment the model is used. But the model could later be updated with additional training (e.g., fine-tuning), which would change its weights. So it's more precise to say that the model is fixed with respect to the moment in which it is used, rather than to take this as a general statement about a model being permanently fixed. *See Talkin'*, *supra* note 20, at 287.

43. *See* Cooper & Grimmelmann, *supra* note 9, at 193–94. This is true in principle, but may vary in practice. For instance, it's true for a given loaded model run with a particular generation configuration on particular hardware. But loading the model weights in a different numerical precision, using generation with or without a KV cache, or other runtime differences may have subtle impacts on the distribution that change which token is ranked at the top. At the system level, there are additional considerations that impact non-determinism of model outputs. For instance, many deployed models are mixtures of experts (MoEs)—there are actually many models, as opposed to a single one. Different prompts get routed to different experts, and the routing decisions may depend on a variety of factors, introducing non-determinism. Nevertheless, for an open-weight model being used with a deterministic decoding algorithm on specific hardware, this is generally true. The results for extracting effectively a verbatim copy of *Harry Potter and the Sorcerer's Stone* using Llama 3.1 70B are deterministic, for the different hardware we ran our code on: with the same prompt, you get the same output. See *infra* Section II.B, discussing *Extracting Books (Open-Weight)*, *supra* note 12. *See generally* A. Feder Cooper, Jonathan Frankle & Christopher De Sa, *Non-Determinism and the Lawlessness of Machine Learning Code* (discussing how hardware and other computational realities impact non-determinism), *in* PROCEEDINGS OF THE 2022 SYMPOSIUM ON COMPUTER SCIENCE AND LAW 1 (Ass'n for Computing Mach. 2022).

determines which of the (typically) many[44] possible continuations represented by the model are realized in any particular output.[45]

Seen in this way, it becomes clear why analogies between models and traditional databases can be misleading. Language models don't store responses in a discrete, localized way and then deterministically return those responses.[46] Instead, they generate outputs from a probability distribution shaped by the training data, with decoding determining which continuation is actually produced in response to any given prompt. These two components—model weights and decoding procedures—together produce content in response to an initial prompt. This distinction is essential for understanding memorization and its relationship to model outputs.[47]

One final clarification is worth noting. The discussion above concerns language models in isolation—that is, the model as defined by its weights and the probability distribution those weights induce over token sequences. This is a useful description for understanding open-weight models, such as Meta's Llama models[48] and Alibaba's Qwen models,[49] which can be downloaded and used directly with a user's chosen decoding procedure.

In practice, however, many end users interact with language models that are embedded within larger proprietary systems: software services deployed as products like Claude or ChatGPT. These systems generally include additional components that constrain or otherwise shape the outputs returned to users such as prompt processing or output filtering. As a result, for these systems, decoding is not the only mechanism that determines outputs; system-level features also play a significant role.[50] Those features may mean that a model actually deployed to the public may not return a particular output to end users even though the model has memorized the content of that output.

44. Memorization, as we will see soon, is an exception to this: for memorized sequences, there are *not* always "many possible continuations" for reasonable decoding procedures; rather, there may effectively only be *one* or a handful of possible continuation. *See infra* Section II.B.

45. *See* Cooper & Grimmelmann, *supra* note 9, at 151–52, 151 nn. 32–33, 152 nn.35–38.

46. *See id.* at 189–90.

47. *See infra* Section II.B. Memorization is compressed information—it's compressed storage—of specific training data in the model; extraction effectively reads those compressed stored data out of the model as output.

48. *See generally* Aaron Grattafiori et al., The Llama 3 Herd of Models (Nov. 23, 2024) (unpublished manuscript), https://arxiv.org/pdf/2407.21783 [https://perma.cc/3H2V-M3BD].

49. *See generally* An Yang et al., Qwen2.5 Technical Report (Jan. 3, 2025) (unpublished manuscript), https://arxiv.org/pdf/2412.15115 [https://perma.cc/J8EN-HNEP].

50. For a significantly more detailed treatment on this topic, we refer to Cooper & Grimmelmann, *supra* note 9. Here, we try to pare things down to what we consider the absolute minimum information needed to dispel misunderstandings of metaphors about "memorization" and "compression." *See infra* Section II.C.

### *B. Memorization in the Model, Extraction in Model Outputs*

Following from above, we can now explain what the machine learning literature refers to as "memorization."[51] In the discussion that follows, a key takeaway is that, when it comes to memorization, it isn't possible to make broad generalizations about AI, or even generalizations about a specific AI model with respect to all the training data it's been trained on.

As detailed above, a trained language model defines a probability distribution over token sequences, where that distribution is learned from the training data. The technical notion of memorization can be understood as a property of that distribution: loosely speaking, memorization is when, based on its training, the model places unusually high probability on specific sequences from its training data. Thus, those sequences are far more likely to appear in the output than alternatives. This doesn't mean the model ceases to be probabilistic; it still defines a probability distribution and may still give non-zero probability to many possible sequences. But the distribution is so sharply peaked that one sequence dominates.[52]

This is what is meant, in a technical sense, when we and others in the machine learning community say that memorized training data are encoded in the model.[53] In that same technical sense, the point is closely related to compression, which helps make the claim more precise. As noted above, learning involves compression: a high-quality trained model stores a compact internal representation of the data that is sufficient for accurate prediction.[54] That compressed representation may capture broad patterns or abstractions, rather than preserving much that is specific to any particular training sequence. This is the intuition behind Ted Chiang's description of ChatGPT as a "blurry JPEG of

51. To understand "sharply peaked": imagine rolling a weighted die, where there's a 95% chance of rolling 3, and the remaining 5% is shared among 1, 2, 4, 5, and 6. This intuition, while useful, is also a simplification in several respects. Scientific research shows that memorized sequences need not be verbatim copies of training data, but may instead be very close variants (e.g., the memorized sequence replaces a comma with a semi-colon). In fact, many near-exact variants may be memorized simultaneously: a model may concentrate probability on a set of such near-verbatim sequences, for example, sequences that differ only in punctuation or exhibit minor syntactic variation. This is still memorization. More broadly, the scientific understanding of memorization continues to evolve. The account here is intended as a high-level description of the aspects that are relevant for copyright. *See* Daphne Ippolito, Florian Tramèr, Milad Nasr, Chiyuan Zhang, Matthew Jagielski, Katherine Lee, Christopher A. Choquette Choo & Nicholas Carlini, *Preventing Generation of Verbatim Memorization in Language Models Gives a False Sense of Privacy* (arguing that "verbatim memorization definitions are too restrictive and fail to capture more subtle forms of memorization"), *in* PROCEEDINGS OF THE 16TH INTERNATIONAL LANGUAGE GENERATION CONFERENCE 28, 28 (C. Maria Keet, Hung-Yi Lee & Sina Zarrieß eds., 2023). *See generally Near-Verbatim*, *supra* note 13 (developing new search models to understand near-verbatim extraction risk).

52. *See, e.g.*, *Extracting Training Data from LLMs*, *supra* note 11; *Extracting Training Data from Diffusion Models*, *supra* note 15; Nasr et al., *supra* note 12; *Carlini Blog*, *supra* note 15; *Extracting Books (Open-Weight)*, *supra* note 12; *Near-Verbatim*, *supra* note 13; Franceschelli et al., *supra* note 15; Cooper & Grimmelmann, *supra* note 9.

53. *See* Delétang et al., *supra* note 29; First Amended Complaint, *supra* note 36, at 27.

54. *See* Delétang et al., *supra* note 29, at 1–3.

all the text on the Web."[55] But memorization reveals the limits of that analogy: in certain cases, compression preserves particular training data with high fidelity. As Cooper & Grimmelmann write, it is perhaps best to

> . . . refine Chiang's blurry JPEG compression analogy: not all parts of the JPEG are equally or completely blurry. Some training data are compressed more than others, and compression happens in different ways. Some of the compression is lossy: the information is discarded or transformed. But the portions that contain exact memorization are lossless [compression] and the portions that contain near-exact memorization are nearly lossless [compression.][56]

So far, we haven't said much about model outputs; what we've just described—memorization—is a property of the model itself. This property is not directly observable. As discussed in Section II.A, the internal structure of models is difficult to interpret, and so we typically can't identify where or how particular information is stored, including memorized training data. As a result, memorization in generative AI is usually studied through the outputs that a model produces under carefully chosen conditions. This isn't because outputs define memorization, but because they provide a way of observing a property of the model that would otherwise remain hidden.

*Figure 2: Four portions of the text of Harry Potter and the Sorcerer's Stone, generated by the authors using Llama 3.1 70B. Highlighting shows differences between the generation and ground-truth text from Books3. Black text is identical across both texts. Text crossed out in red indicates ground-truth text from the book absent in the generation. Text highlighted in yellow is present in the generation, but absent from the ground-truth book. Note that this visual diffing procedure is sensitive to benign formatting differences. These four samples highlight different types of variations we observe between the ground-truth book and the generation. The large majority are formatting differences and British-versus-American English localization changes. (We have the British version available to us from Books3, but Llama 3.1 70B generated the book with American spelling.) Overall, simple quantitative metrics comparing syntactic similarity for the generated text and the ground-truth book yield a 99.2% match. This is why we say that our results produced a near-exact copy of* the entire book.

55. Ted Chiang, *ChatGPT is a Blurry JPEG of the Web*, NEW YORKER (Feb. 9, 2023), https://www.newyorker.com/tech/annals-of-technology/chatgpt-is-a-blurry-jpeg-of-the-web.

56. *See* Cooper & Grimmelmann, *supra* note 9, at 188–89. We discuss the difference between lossy and lossless compression *infra* notes 101–03 and accompanying text.

Images drawn from Cooper et al. Extracting memorized pieces of (copyrighted) books from open-weight language models; reprinted with permission.

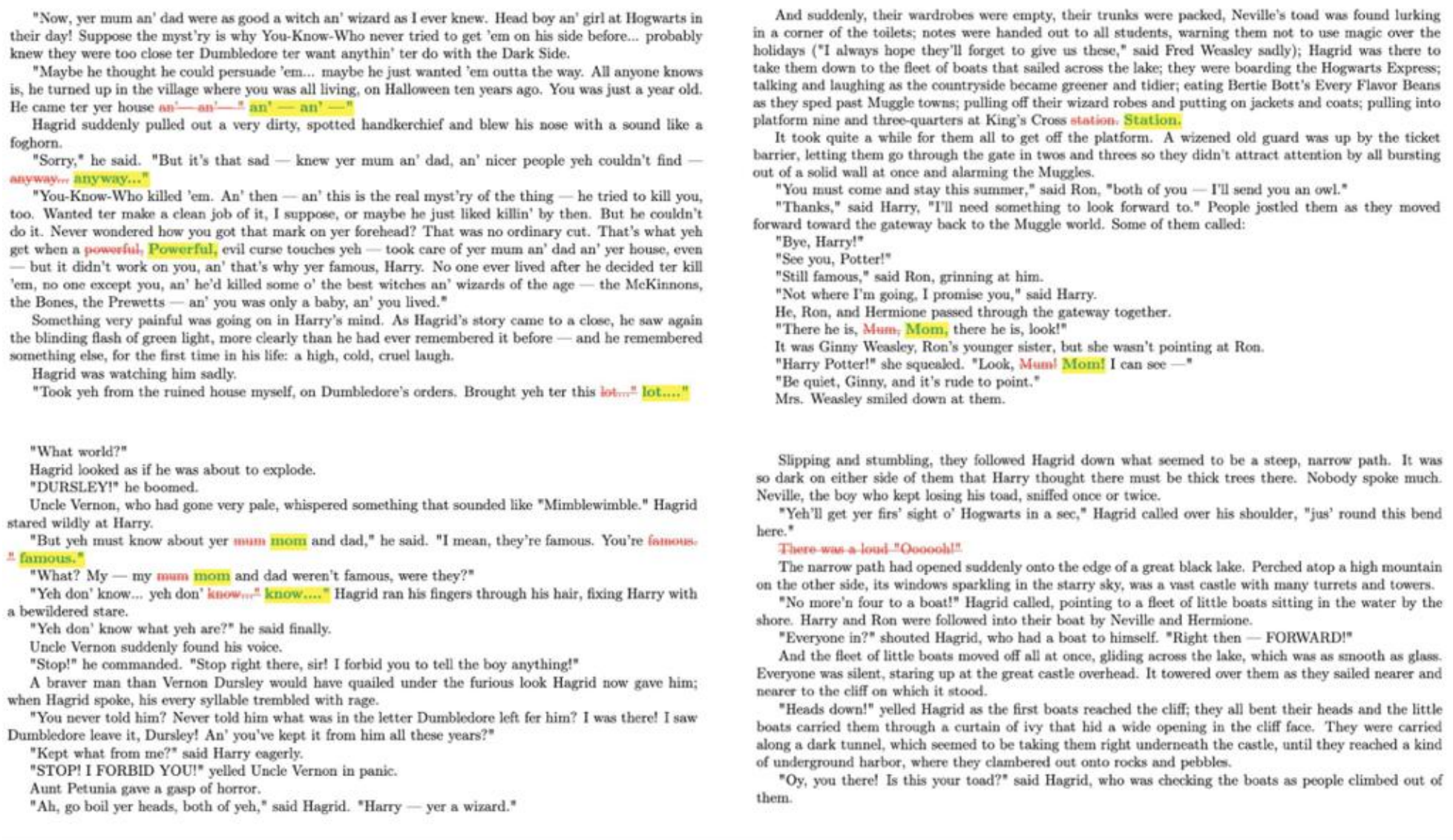

"Now, yer mum an' dad were as good a witch an' wizard as I ever knew. Head boy an' girl at Hogwarts in their day! Suppose the myst'ry is why You-Know-Who never tried to get 'em on his side before... probably knew they were too close ter Dumbledore ter want anythin' ter do with the Dark Side.

"Maybe he thought he could persuade 'em... maybe he just wanted 'em outta the way. All anyone knows is, he turned up in the village where you was all living, on Halloween ten years ago. You was just a year old. He came ter yer house ~~an'—an'—"~~ an' — an' —"

Hagrid suddenly pulled out a very dirty, spotted handkerchief and blew his nose with a sound like a foghorn.

"Sorry," he said. "But it's that sad — knew yer mum an' dad, an' nicer people yeh couldn't find — ~~anyway...~~ anyway..."

"You-Know-Who killed 'em. An' then — an' this is the real myst'ry of the thing — he tried to kill you, too. Wanted ter make a clean job of it, I suppose, or maybe he just liked killin' by then. But he couldn't do it. Never wondered how you got that mark on yer forehead? That was no ordinary cut. That's what yeh get when a ~~powerful,~~ Powerful, evil curse touches yeh — took care of yer mum an' dad an' yer house, even — but it didn't work on you, an' that's why yer famous, Harry. No one ever lived after he decided ter kill 'em, no one except you, an' he'd killed some o' the best witches an' wizards of the age — the McKinnons, the Bones, the Prewetts — an' you was only a baby, an' you lived."

Something very painful was going on in Harry's mind. As Hagrid's story came to a close, he saw again the blinding flash of green light, more clearly than he had ever remembered it before — and he remembered something else, for the first time in his life: a high, cold, cruel laugh.

Hagrid was watching him sadly.

"Took yeh from the ruined house myself, on Dumbledore's orders. Brought yeh ter this ~~lot..."~~ lot...."

"What world?"

Hagrid looked as if he was about to explode.

"DURSLEY!" he boomed.

Uncle Vernon, who had gone very pale, whispered something that sounded like "Mimblewimble." Hagrid stared wildly at Harry.

"But yeh must know about yer ~~mum~~ mom and dad," he said. "I mean, they're famous. You're ~~famous."~~ famous."

"What? My — my ~~mum~~ mom and dad weren't famous, were they?"

"Yeh don' know... yeh don' ~~know..."~~ know...." Hagrid ran his fingers through his hair, fixing Harry with a bewildered stare.

"Yeh don' know what yeh are?" he said finally.

Uncle Vernon suddenly found his voice.

"Stop!" he commanded. "Stop right there, sir! I forbid you to tell the boy anything!"

A braver man than Vernon Dursley would have quailed under the furious look Hagrid now gave him; when Hagrid spoke, his every syllable trembled with rage.

"You never told him? Never told him what was in the letter Dumbledore left fer him? I was there! I saw Dumbledore leave it, Dursley! An' you've kept it from him all these years?"

"Kept what from me?" said Harry eagerly.

"STOP! I FORBID YOU!" yelled Uncle Vernon in panic.

Aunt Petunia gave a gasp of horror.

"Ah, go boil yer heads, both of yeh," said Hagrid. "Harry — yer a wizard."

And suddenly, their wardrobes were empty, their trunks were packed, Neville's toad was found lurking in a corner of the toilets; notes were handed out to all students, warning them not to use magic over the holidays ("I always hope they'll forget to give us these," said Fred Weasley sadly); Hagrid was there to take them down to the fleet of boats that sailed across the lake; they were boarding the Hogwarts Express; talking and laughing as the countryside became greener and tidier; eating Bertie Bott's Every Flavor Beans as they sped past Muggle towns; pulling off their wizard robes and putting on jackets and coats; pulling into platform nine and three-quarters at King's Cross ~~station.~~ Station.

It took quite a while for them all to get off the platform. A wizened old guard was up by the ticket barrier, letting them go through the gate in twos and threes so they didn't attract attention by all bursting out of a solid wall at once and alarming the Muggles.

"You must come and stay this summer," said Ron, "both of you — I'll send you an owl."

"Thanks," said Harry, "I'll need something to look forward to." People jostled them as they moved forward toward the gateway back to the Muggle world. Some of them called:

"Bye, Harry!"

"See you, Potter!"

"Still famous," said Ron, grinning at him.

"Not where I'm going, I promise you," said Harry.

He, Ron, and Hermione passed through the gateway together.

"There he is, ~~Mum,~~ Mom, there he is, look!"

It was Ginny Weasley, Ron's younger sister, but she wasn't pointing at Ron.

"Harry Potter!" she squealed. "Look, ~~Mum!~~ Mom! I can see —"

"Be quiet, Ginny, and it's rude to point."

Mrs. Weasley smiled down at them.

Slipping and stumbling, they followed Hagrid down what seemed to be a steep, narrow path. It was so dark on either side of them that Harry thought there must be thick trees there. Nobody spoke much. Neville, the boy who kept losing his toad, sniffed once or twice.

"Yeh'll get yer firs' sight o' Hogwarts in a sec," Hagrid called over his shoulder, "jus' round this bend here."

~~There was a loud "Oooooh!"~~

The narrow path had opened suddenly onto the edge of a great black lake. Perched atop a high mountain on the other side, its windows sparkling in the starry sky, was a vast castle with many turrets and towers.

"No more'n four to a boat!" Hagrid called, pointing to a fleet of little boats sitting in the water by the shore. Harry and Ron were followed into their boat by Neville and Hermione.

"Everyone in?" shouted Hagrid, who had a boat to himself. "Right then — FORWARD!"

And the fleet of little boats moved off all at once, gliding across the lake, which was as smooth as glass. Everyone was silent, staring up at the great castle overhead. It towered over them as they sailed nearer and nearer to the cliff on which it stood.

"Heads down!" yelled Hagrid as the first boats reached the cliff; they all bent their heads and the little boats carried them through a curtain of ivy that hid a wide opening in the cliff face. They were carried along a dark tunnel, which seemed to be taking them right underneath the castle, until they reached a kind of underground harbor, where they clambered out onto rocks and pebbles.

"Oy, you there! Is this your toad?" said Hagrid, who was checking the boats as people climbed out of them.

To make this concrete, we turn to a notable example of memorization identified in our own prior work.[57] By studying the outputs of open-weight, non-chatbot[58] models—models that are widely available for download—we found that these models have memorized significantly more text from books in their training data than was previously understood. In an extreme instance, prompting Meta's Llama 3.1 70B model with just a short snippet from the opening of *Harry Potter and the Sorcerer's Stone* was sufficient to generate a near-exact copy of the entire 304-page book. For behavior like this, the only reasonable explanation is memorization of the book from the training data.[59] Figure 2 shows what we mean by near-exact: for four portions of the generated text, it shows the

57. *See Extracting Books (Open-Weight)*, *supra* note 12, at 13–15.

58. These are models that have not been fine-tuned, aligned, or otherwise post-trained to behave like conversational chatbots. See *Talkin'*, *supra* note 20, at 286–88, laying out the generative AI supply chain, including model training, fine tuning, and alignment, and *id.* at 306–307, for a more detailed discussion of model alignment. It's possible that models are aligned or fine-tuned in other ways than to make them chatbot-like. This makes the model different from the original base model from pre-training —it's no longer a base model—even though it isn't a chatbot. This complexity is why we inelegantly refer to the type of model we're talking about as a "non-chatbot model." For more on this regarding open-weight models like Llama 2 70B (which is base model) and Llama 3.1 70B (which is not, because it underwent some post-training, even though it's not a chatbot), see *Extracting Books (Open-Weight)*, *supra* note 12, at 6–8. These models are both different from their "Instruct" variants (e.g., Llama 2 70B Instruct, Llama 3.1 70B Instruct), which are chatbots derived from their respective non-chatbot models. All of these are LLMs, and they're all available for commercial use.

59. For an exhaustive treatment of this, see *Extracting Books (Open-Weight)*, *supra* note 12, at 163–81; *see also Carlini Blog*, *supra* note 15; Cooper & Grimmelmann, *supra* note 9.

differences between the generated text and the reference, ground-truth text we have available from the Books3 corpus.[60] Using the same code with Llama 3.1 70B and a prompt of only the first sentence, we also generated hundreds of pages of George Orwell's *Nineteen Eighty-Four* near-verbatim.

This behavior can be understood in terms of the probability distribution defined by the model. Starting from this very short prompt, the next tokens in the original text have overwhelmingly higher probability than plausible alternatives. Figure 1 provides a simplified illustration for *Harry Potter and the Sorcerer's Stone*, showing the prompt, the model's highest-probability tokens at each step, and the token actually selected during decoding. The ground-truth text from the book consistently has by far the highest probability. The distribution is so sharply peaked that, in practice, the generated output tracks the memorized text. For Llama 3.1 70B, this is true throughout *Harry Potter and the Sorcerer's Stone*,[61] such that it is possible to deterministically reproduce the entire book. This type of behavior—the generation of memorized training data in model outputs—is commonly referred to in the machine learning literature as "extraction."[62]

*Figure 3: Running the same long-form extraction procedure on Sandman Slim. The blue text is the prompt we provided, which is the first few sentences of Richard Kadrey's book. The black text is the generation. We tried prompts of various lengths, but produced no generations that were near-verbatim reproductions of text in the ground-truth book. This experiment doesn't show successful extraction of memorized training data. (Interestingly, though, the*

60. The Books3 corpus contains approximately 200,000 torrented ebooks, converted to text-format for the purpose of LLM training. Alex Reisner, *Revealed: The Authors Whose Pirated Books are Powering Generative AI*, THE ATLANTIC (Sep. 25, 2023, at 13:40 ET), https://www.theatlantic.com/technology/archive/2023/08/books3-ai-meta-llama-pirated-books/675063/ (on file with the Berkeley Technology Law Journal); *see Talkin'*, *supra* note 20, at 37–39, 113–14. *See generally* Katherine Lee, A. Feder Cooper, James Grimmelmann & Daphne Ippolito, AI and Law: The Next Generation (Oct. 19, 2023) (unpublished manuscript), https://papers.ssrn.com/sol3/papers.cfm?abstract_id=4580739 (on file with the Berkeley Technology Law Journal) (describing pre-training datasets, including Books3).

61. The probabilities are so high that we are able to deterministically decode the whole book using a decoding algorithm called beam search. *See Extracting Books (Open-Weight)*, *supra* note 12, at 1, 183.

62. *See generally* Cooper & Grimmelmann, *supra* note 9 (giving a more precise taxonomy of terms and unpacking some distinctions in the literature). Extraction is intentional prompting to attempt to reproduce training data in outputs. Regurgitation is the unintentional reproduction of training data in outputs. The frequency of regurgitation in language models is poorly understood and not often studied in the literature. There are two exceptions (only sort of) in very limited and controlled research settings. *See generally* Michael Aerni, Javier Rando, Edoardo Debenedetti, Nicholas Carlini, Daphne Ippolito & Florian Tramèr, *Measuring Non-Adversarial Reproduction of Training Data in Large Language Models*, THIRTEENTH INT'L CONF. ON LEARNING REPRESENTATIONS, Feb. 11, 2025, art. 11752; Parjanya Prajakta Prashant, Kaustubh Ponkshe & Babak Salimi, *TokenSwap: A Lightweight Method to Disrupt Memorized Sequences in LLMs*, THIRTY-NINTH ANN. CONF. ON NEURAL INFO. PROCESSING SYS., Sep. 18, 2025, art. 21464. We defer to Cooper & Grimmelmann, *supra* note 9, for more precise treatment, but will call this extraction here.

*generation does contain references to characters in the book, which we did not provide in the prompt: James Stark, Sandman Slim, etc.)*

I WAKE UP on a pile of smoldering garbage and leaves in the old Hollywood Forever cemetery behind the Paramount Studio lot on Melrose, though these last details don't come to me until later. Right now all I know is that I'm back in the world and I'm on fire. My mind hasn't quite kicked in yet, but my body knows enough to roll off the burning trash and to keep rolling until I can't feel the heat anymore. I come to a stop against a gravestone and lie there for a moment, trying to remember how I got here and who I am. I don't always know right away. It's the amnesia. It's a hell of a way to travel, but it's better than being dead. Slowly, like dirty water dropping out of a rusty pipe, it comes back to me. My name is James Stark. Friends call me Sandman Slim. Call me what you like. I've been called worse. I got this name because all I do is bring nightmares. True, I spend a lot of time in the City of Angels, but it's not like they're putting up a statue to me. Before this, I lived in Hell for eleven years. Even when I tried to stay out of trouble, trouble would come right to my door and make friends. That kind of thing has a way of making a man a little jaded when it comes to trusting his fellow man. Or demons. Even angels. I'm like that. I've got a bad attitude. I drink. I've been kicked out of every twelve-step program I've ever joined. I don't have many friends, so the ones I have mean a lot to me. They mean enough that if someone hurts them, I'll tear the universe down to save them. I'm telling you this so you'll know who I am and what I'm talking about when I say there are things in the dark. Things we don't always see that move among us. Some of these things are good and some aren't and sometimes they do things that make you want to rip open your chest and pull out your heart because you've seen too much and there's no going back. I know about these things because I'm one of them. I'm a monster and I'm a killer. I'm good at it and I never let emotion get in the way. When you live where I live, love can get you killed. And when it doesn't, it's worse because if you don't die, you just wish

Extraction isn't always possible. Models frequently generate text that doesn't reproduce memorized training data. As a counterpoint to the Harry Potter example, consider our attempt to apply a similar procedure for another Kadrey v. Meta book, Richard Kadrey's Sandman Slim, which we illustrate in Figure 3. We know that this book was in Llama's training data,63 so it's at least possible in principle for the model to have memorized text from it, just as it memorized the first Harry Potter book or Nineteen Eighty-Four. But in practice, using the same procedure, the model doesn't reproduce the book. Instead, it produces a range of plausible outputs, reflecting a distribution that isn't concentrated on any single sequence, let alone the ground-truth text from the training data.

These contrasting examples illustrate an important point: memorization isn't a uniform phenomenon that applies equally across a model's training data. Some training data are memorized, while others aren't. Extraction is an imperfect tool for measuring memorization:[64] it can, by definition, only produce sequences that have been memorized,[65] and not all memorized training data are necessarily

63. Kadrey v. Meta Platforms, Inc., 788 F. Supp. 3d 1026, 1042 (N.D. Cal. 2025).

64. Specific extraction procedures are selective in additional ways. For instance, a given procedure may test only a certain set of prompts, and miss identifying memorization that other prompts would have revealed. Research using greedy decoding misses large amounts of extraction that ours and others' work has identified using probabilistic extraction, and so on. See *Extracting Books (Open-Weight)*, *supra* note 12, at 186–88; *Near-Verbatim*, *supra* note 13, at 8–10.

65. An important consideration here is that the measurement procedure for extraction has to be *valid*. If it isn't valid, then the "extraction" procedure could produce outputs that are not actually extraction, but chance generation of *non*-training data that would erroneously get interpreted as evidence of memorization. For rigorous scientific work on extraction and memorization, this is not an issue. Nevertheless, not all published work in the technical literature meets this bar. The invalidity of some work that calls itself "extraction" does not cut against the validity of work that rigorously studies extraction. However, it's important to be vigilant about this when assessing extraction claims. This of course isn't unique to extraction; distinguishing between valid and invalid measurement methodologies extractable.[66] In this second sense, extraction is an imperfect measurement tool: it provides evidence of memorization where it succeeds, but its failure doesn't necessarily imply that memorization hasn't occurred. Memorization isn't a Schrödinger's-cat-like situation: memorized training data do not both exist and not exist depending on whether they're observed. Training data may or may not be encoded in the model; and if they are, they are there regardless of whether they can be extracted and whether we can prove it.[67]

Further, models don't all compress information about their training data in the same way. Sometimes they capture high-level relationships across data points without memorizing specific text sequences, and sometimes they encode specific sequences (i.e., memorize them). Sometimes memorized training data can be extracted deterministically, and sometimes only probabilistically, with varying frequency.[68] These details—the probabilistic elements, how compression works, etc.—are the core of the issue and can't be abstracted away.[69]

But even with this clarifying perspective, much about memorization remains poorly understood. Some factors, such as duplication of sequences in training data, are known to increase the likelihood of memorization.[70] However, in general, there is no single answer to how much memorization occurs, or precisely when and why it arises, in part because our tools for measuring it are imprecise at best. For instance, some prior estimates suggest that somewhere between 0.1% and 10% of training data sequences may be memorized by language models.[71] While these numbers might sound like firm, general, authoritative claims, they are in fact based on very specific experimental and resulting claims is a challenge for empirical sciences, more generally. *See generally* A. Feder Cooper, Marika Swanberg, Jamie Hayes, Lea Duesterwald, Christopher De Sa, Daniel E. Ho, Mark A. Lemley & Percy Liang, Extractable Memorization from First Principles (July 14, 2026) (unpublished manuscript) [hereinafter First Principles], https://arxiv.org/pdf/2607.12649 [https://perma.cc/9C7F-KTJR].

66. Cooper & Grimmelmann, *supra* note 9, at 177–78; Nasr et al., *supra* note 12, at 1 (distinguishing discoverable and extractable memorization); *see* First Principles, *supra* note 65, at 4.

67. See *infra* Section II.C for a discussion of how this point is frequently misunderstood: extraction doesn't determine memorization (this flips the relevant direction); memorization is a prerequisite for extraction.

68. *See* Jamie Hayes, Marika Swanberg, Harsh Chaudhari, Itay Yona, Ilia Shumailov, Milad Nasr, Christopher A. Choquette-Choo, Katherine Lee & A. Feder Cooper, *Measuring Memorization in Language Models via Probabilistic Extraction*, *in* 1 PROCEEDINGS OF THE 2025 CONFERENCE OF THE NATIONS OF THE AMERICAS CHAPTER OF THE ASSOCIATION FOR COMPUTATIONAL LINGUISTICS: HUMAN LANGUAGE TECHNOLOGIES 9266 (NAACL 2025). *See generally Near-Verbatim*, *supra* note 13.

69. *See* Cooper & Grimmelmann, *supra* note 9, at 197 ("[T]he probabilistic element . . . is inescapable.").

70. *Deduplicating*, *supra* note 11, at 7.

71. *See* Cooper & Grimmelmann, *supra* note 9, at 201–04; Nasr et al., *supra* note 12, at 15 ("[T]his means that approximately 10% of the entire model capacity is 'wasted' on verbatim memorized training data.").

setups.[72] They are average estimates computed using specific extraction procedures on particular models and datasets.

Changing any of these factors—the extraction method, the model, or the training data of interest—can significantly affect the results. Much of the early work relied on greedy decoding to measure extraction, which selects the highest-probability token at each decoding step.[73] More recent approaches allow for more realistic non-deterministic decoding schemes and repeated sampling from the same prompt, which can reveal additional instances of memorization that greedy methods miss.[74] In our own work, we quantify how running the same prompt hundreds or thousands of times can produce verbatim extraction of copyrighted text in some runs,[75] but for many works, a single verbatim extraction of fifty tokens from a work might be accompanied by hundreds of outputs that are not verbatim extractions.[76] And our more recent work shows that even these newer approaches undercount the extent of memorization, because many of the "failures" to extract a sequence in fact generate near-verbatim copies of that sequence.[77]

What we *can* say with some confidence from the empirical extraction literature is that memorization doesn't reduce to a simple story. It's not correct to conclude either that AI models always memorize specific content or that they almost never do. Instead, the extent and specificity of memorization vary

72. For example, Morris et al. make an aggregate claim that models in the GPT family have a 3.6 bits-per-parameter memorization capacity. *See generally* John X. Morris, Chawin Sitawarin, Chuan Guo, Narine Kokhlikyan, G. Edward Suh, Alexander M. Rush, Kamalika Chaudhuri & Saeed Mahloujifar, How Much Do Language Models Memorize? (June 18, 2025) (unpublished manuscript), https://arxiv.org/pdf/2505.24832 [https://perma.cc/3ZGD-8RRQ]. This result is derived with respect to a very specific experimental training conditions for small models—a setup that is significantly different from those used to train production LLMs. *See id.* at 2. So while this work offers a precise scientific claim, in isolation, it is easy to misconstrue the "3.6 bits" number as a general claim about LLMs, as in this news coverage of the scientific paper: Carl Franzen, *How Much Information Do LLMs Really Memorize? Now We Know, Thanks to Meta, Google, Nvidia and Cornell*, VENTUREBEAT (June 5, 2025, at 08:35 PT), https://venturebeat.com/technology/how-much-information-do-llms-really-memorize-now-we-know-thanks-to-meta-google-nvidia-and-cornell [https://perma.cc/F8ML-TN9K]. *See generally* First Principles, *supra* note 65 (concerning how the practical setup for extraction experiments influences conclusions about memorization).

73. *See supra* Section II.A (defining greedy decoding); *Deduplicating*, *supra* note 11, at 5–6; Petko Georgiev et al., Gemini 1.5: Unlocking Multimodal Understanding Across Millions of Tokens of Context 1–2 (Dec. 16, 2024) (unpublished manuscript), https://arxiv.org/pdf/2403.05530 [https://perma.cc/WS3W-5AEJ]; Grattafiori et al., *supra* note 48, at 3; Morgane Riviere et al., Gemma 2: Improving Open Language Models at a Practical Size (Oct. 2, 2024) (unpublished manuscript), https://arxiv.org/pdf/2408.00118 [https://perma.cc/8DGG-2BKY].

74. *See generally* Hayes et al., *supra* note 68.

75. This work validates that it's actually capturing extraction, as opposed to chance generation that resembles training data. We defer the details to that work, but emphasize here that generating training data one out of every 1,000 attempts with the same prompt is strong evidence for memorization in the work we conduct. *See Extracting Books (Open-Weight)*, *supra* note 12, at 11; *see also* First Principles, *supra* note 65, at 12–14 (establishing standards for evaluating claims of memorization).

76. *See Near-Verbatim*, *supra* note 13, at 18–19.

77. *Id.*

significantly across models (e.g., Llama versus Qwen), across versions within the same model family (e.g., Llama 2 versus Llama 3), across model sizes (e.g., Llama 2 7B versus Llama 2 70B), and across individual works for a given model.[78] And for models that are part of deployed AI systems, additional system-level factors further affect whether memorized training data are observed in outputs.[79] As recent work shows, it's definitely possible to extract some copyrighted training data, but, depending on the system, this can require different methods altogether.[80] So, in practice, when it comes to memorization it isn't possible to make broad generalizations about AI, or even generalizations about a specific AI model with respect to all the training data it's been trained on.

### C. *Metaphors, Misunderstandings, and the Problem of Dismissing Technical Precision*

Having gone beneath the metaphorical surface of terms like memorization and compression, it should now be clearer why those terms are not exhausted by their lay interpretations.[81] In machine learning, they name technical phenomena. The mistake, then, is not merely to notice that the metaphors are imperfect—they very much are. It is to treat that imperfection as a reason to discount or abstract away the technical particulars.

A model is not literally arranged like a filesystem or a database. But that doesn't really matter,[82] and that also doesn't mean it stores nothing.[83] It stores

78. Nicholas Carlini, Daphne Ippolito, Matthew Jagielski, Katherine Lee, Florian Tramer & Chiyuan Zhang, Quantifying Memorization Across Neural Language Models, ELEVENTH INT'L CONF. ON LEARNING REPRESENTATIONS, Jan. 14, 2026, at 1; Hayes et al., *supra* note 68; *Near-Verbatim*, *supra* note 13 (discussing probabilistic extraction and near-verbatim probabilistic extraction, for the same point). *See generally Extracting Books (Open-Weight)*, *supra* note 12 (providing a books-specific variation in memorization claims, across books, models, model sizes, and versions).

79. *See* Cooper & Grimmelmann, *supra* note 9, at 215–17 (discussing system-level modifications that can prevent memorized training data from being produced to the user even if it is generated by the model); *Machine Unlearning*, *supra* note 29, at 5.

80. *See* Nasr et al., *supra* note 12, at 9 (causing a "divergence" jailbreak by asking the model to repeat the same word forever, e.g., "poem," resulted in some outputs leaking verbatim training data from ChatGPT 3.5). To extract copyrighted books from proprietary models, Ahmed et al. used jailbreaks to elicit memorization from ChatGPT 4.1 and Claude 3.7 Sonnet, but did not need to jailbreak Grok 3 or Gemini 2.5 Pro to elicit enormous amounts of copyrighted material. *See* Extracting Books (Production LLMs), *supra* note 13, at 1. In all cases, they were able to bypass content filters to reproduce copyrighted material in system outputs). *See id.* at 5–7.

81. *See supra* Section II.A.

82. See *infra* Section III.A and *infra* note 97.

83. *See generally* Aline Larroyed, The Fallacy of the File: How the Memorisation Metaphor Misguides Copyright Law and Stifles AI Innovation (Nov. 21, 2025) (unpublished manuscript), https://papers.ssrn.com/sol3/papers.cfm?abstract_id=5782882 (on file with the Berkeley Technology Law Journal) (conflating the technical details of memorization with the metaphorical aspects of the term memorization, taking the metaphor in the title of Cooper & Grimmelmann's *The Files Are in the Computer: On Copyright, Memorization, and Generative AI* too literally, taking the term "memorisation" only on metaphorical rather than technical terms, and then arguing that because memorization is both metaphorical and not exactly like file storage, it's irrelevant to copyright).

information, in bits, in a technically meaningful sense.[84] In the case of memorized training data that can be extracted, information about those data is encoded in the model's weights with enough fidelity to make those data recoverable in outputs. Extraction is important for precisely that reason: it's evidence of training data that are encoded in the model's weights; it "is a symptom of memorization in the model, not its cause."[85] One therefore can't take extraction seriously while setting aside the memorization in the model, because the former is only possible because of the latter.

It's also true that training a model isn't literally like creating a ZIP file or converting a song into MP3 format, and so thinking about learning and compression in those terms can be misleading. But that observation addresses only the everyday analogy and the possible misuse courts or litigants might make of it; it doesn't address the underlying technical claim.[86] In cases of memorization, compression preserves certain specific training data with high fidelity. For that reason, compression helps explain what has been retained in the model and why some memorized training data may later be more readily recoverable than others.[87]

---

84. *See generally* Delétang et al., *supra* note 29; Zhiying Jiang, Matthew Yang, Mikhail Tsirlin, Raphael Tang, Yiqin Dai & Jimmy Lin, *"Low Resource" Text Classification: A Parameter-Free Classification Method with Compressors*, *in* FINDINGS OF THE ASSOCIATION FOR COMPUTATIONAL LINGUISTICS: ACL 2023, at 6810 (Anna Rogers, Jordan Boyd-Graber & Naoaki Okazaki eds., 2023); THOMAS M. COVER & JOY A. THOMAS, ELEMENTS OF INFORMATION THEORY 103–58 (2d ed. 2006).

85. Cooper & Grimmelmann, *supra* note 9, at 142. See *supra* Section II.B for this point; *see also* Adi Haviv, Niva Elkin-Koren, Uri Hacohen, Roi Livni & Shay Moran, We Should Separate Memorization from Copyright 4–5 (Feb. 9, 2026) (unpublished manuscript), https://arxiv.org/pdf/2602.08632 [https://perma.cc/6LV5-565A]. Haviv et al. emphasize that model weights aren't directly interpretable and that training data aren't stored in a database-like or file-like format. Haviv et al., *supra*, at 5. Then, from that basis, they cast extraction as too mediated by prompts and prior knowledge to be evidence of memorization encoded in the model itself. *Id.* They explain that point by analogy to fMRI reconstructions: brain signals may contain information about an image, but the recognizable image emerges only after a separate reconstruction step infers an image-like output from those signals. *Id.* But that analogy reverses a basic causal point. Prompts and prior knowledge bear on the conditions under which memorized training data becomes observable, not on whether those training data were encoded in the model in the first place. A selective or prompt-dependent extraction result shows only that memorized information isn't uniformly decodable from the model's weights; it doesn't show that the information was supplied by the extraction procedure rather than encoded through training. Indeed, properly validated extraction procedures are designed to rule out the prompt itself as the source of the recovered material. *See Extracting Books (Open-Weight)*, *supra* note 12, at 1–2, 4–5. The fMRI comparison is therefore misleading here. An fMRI scanner produces only an indirect physiological trace, and the recognizable image appears only after a separate system maps that trace back into the image space. By contrast, in extraction, there's no separate reconstruction system generating the recovered content: the model being studied generates the sequence itself.

86. See Sag, *supra* note 39, at 38–39, recognizing that there are instances of memorization—the technical model behavior—that are salient to copyright law; but rejecting the "fallacy of compression" on analogical grounds because "compression" invokes things like MP3 files in ordinary meaning. In doing so, the paper doesn't directly engage the technical aspects of compression and memorization that it simultaneously treats as relevant to copyright. *Id.*

87. *See generally* Hayes et al., *supra* note 68; *Near-Verbatim*, *supra* note 13.

The point, then, is not to discard terms like memorization and compression because their everyday meanings are imperfect guides to how generative AI works. It's to engage them on their technical particulars, as we have done here, because doing so reveals details about what's in the model and what may be recovered in outputs—details that matter for legal analysis. Part II turns to those consequences for copyright.

## III.
## THE LEGAL COMPLICATIONS OF A "COPY"

Part I showed that memorization presents several technical uncertainties. New extraction methods continue to reveal memorized material that earlier methods may not expose, making it difficult to say with confidence that a particular piece of training data definitively hasn't been memorized by a particular model. And where extraction is possible, in some cases it may occur only probabilistically, rather than being deterministically guaranteed. To these technical uncertainties, we now add a discussion of legal uncertainty.

It's unclear how the definition of a "copy" applies to the storage of information that can sometimes be used probabilistically to generate identical or substantially similar outputs in response to a prompt. As discussed below, we can't simply rely on the statutory definition of a copy, in part because it is itself internally self-contradictory. And while courts have applied it to a variety of situations, they haven't confronted one quite like this before. How the courts decide to treat stored model weights that can be used to probabilistically generate different outputs,[88] some but not all of which correspond to prior works, will have significant consequences for the potential liability of AI companies, the viability of open-source business models, the feasibility of efforts to minimize output infringement, and the feasibility of licensing efforts.

### *A. The Statute Doesn't Tell Us Whether Model Weights Are a Copy*

Section 101 of the Copyright Act is normally quite a good guide to the meaning of terms. Unfortunately, that's not true here.

We begin with a significant problem, albeit one tangential to our issue: the definition of copy in the Act is self-contradictory.[89] Read literally, nothing can possibly be an infringing copy. Section 106 says "the owner of copyright under this title has the exclusive rights to do and to authorize any of the following: (1)

88. And, in some cases, deterministically the same output. *See supra* Section II.A (defining greedy decoding); *supra* Figure 1.

89. This hasn't really been discussed in the prior literature. Tony Reese and Tyler Ochoa advert to the confusing language in footnotes. *See* R. Anthony Reese, *Public Display Right*, 2001 U. ILL. L. REV. 83, 127 n.171 (2001); Tyler T. Ochoa, *Copyright, Derivative Works and Fixation: Is* Galoob *a Mirage, or Does the* Form(gen) *of the Alleged Derivative Work Matter?*, 20 SANTA CLARA COMPUT. & HIGH TECH. L.J. 991, 997 n.26 (2004).

to reproduce the copyrighted work in copies."[90] So, not surprisingly, one infringes a copyright by making a copy. Section 101 defines "Copies" as "material objects, other than phonorecords, in which a work is fixed."[91] That means that transitory things like projection of a movie on a screen or data that transits a wire or radio waves in electronic form are not infringing copies. But section 101 also says that "a work is 'fixed' in a tangible medium of expression when its embodiment in a copy or phonorecord, *by or under the authority of the author*, is sufficiently permanent or stable to permit it to be perceived, reproduced, or otherwise communicated for a period of more than transitory duration."[92]

Putting those three together and taking only the literal language seriously, the only potentially infringing "copy" is one that is made "by or under the authority of the author"—and therefore is presumably not infringing.

To be clear, the point here is not that a court is likely to conclude that nothing is copyright infringement. Courts have consistently ignored (or perhaps not discovered) this problem in the statute and proceeded to declare infringing acts copies.[93] Rather, our point is that because the statute is incoherent on its face, we ultimately can't rely on plain meaning in interpreting the statute, and may need to turn to policy, logic, and the purpose of the law to understand it.[94]

At a minimum, though, the statute tells us that to be a copy, the copyrighted work in the model must be fixed in a material object. Computers definitely count as material objects,[95] but there must be actual storage of the copyrighted work in the computer, not just the potential for the computer to generate the work.

That storage need not look like the object in the real world. It's well established that a computer that stores a book or a movie in the form of electrical impulses that translate to 1s and 0s still contains a copy. The definition of a copy addresses this directly; a copy is a material object "in which a work is fixed by any method now known or later developed, and from which the work can be perceived, reproduced, or otherwise communicated, either directly or with the aid of a machine or device."[96] So it's certainly possible that a work could be

90. 17 U.S.C. § 106.

91. § 101.

92. *Id.* (emphasis added).

93. The problem arises from the perhaps unfortunate choice to define "copy" as central to both copyrightability and infringement without distinguishing the two. For a discussion, see PAUL GOLDSTEIN, GOLDSTEIN ON COPYRIGHT § 7.1 (3d ed. 2026-2 Supp. 2026).

94. It also is yet another piece in the mountain of evidence that "plain meaning" is a rather foolish doctrine to adhere to blindly. But that's beyond the scope of the paper.

95. Congress amended the Copyright Act in 1980 to add a definition of computer programs, and the legislative history indicates an intent that a computer program be treated as a literary work, but curiously the amended statute nowhere says as much.

96. 17 U.S.C. § 101.

"fixed" in a series of model weights—*if* it can reliably be perceived or reproduced from those model weights.[97]

### *B. The Challenge of Non-Deterministic Instantiations*

The embodiment of the work must actually exist in the model in some semi-permanent form. The statute says the embodiment in a material object must be "sufficiently permanent or stable to permit it to be perceived, reproduced, or otherwise communicated."[98] That turns out to be a key question for a non-deterministic technology like generative AI. In many prior technologies, the embodiment in a computer was encoded, but often deterministic: the same compiled code would create the same result by calling on the same 1s and 0s each time. That deterministic reproduction is what people had in mind when talking about fixation in a computer in the 1976 Copyright Act.[99] If we think the AI is calling on a stored version of the work as a whole, that is clearly satisfied. (The scientific work in machine learning and extraction is consistent with this: there are bits stored in the model that encode memorized works, even if current scientific methods can't localize which model weights reflect that encoding.) If, however, we think the model is generating an output on the fly, it wouldn't be; the user is fixing—and therefore creating—the output by prompting the AI, which creates something new in response to the prompt.[100]

There are some prior technologies that sit somewhere in between those examples. One is "lossy compression." While some computer programs store information with perfect fidelity in a database, often that takes up too much memory. Compression technologies allow a computer to store a smaller subset of information corresponding to, say, a music file or a photograph and then reconstitute the full song or image from the compressed data.[101] Some compression is "lossless": the computer essentially has a shorthand way of storing a reference to all the information in the original file, and it expands that shorthand into the full information set on demand. Other compression is "lossy": we sacrifice some image quality and fidelity in order to store the image in a smaller amount of space. A file saved with lossy compression doesn't correspond exactly to the original work. Nonetheless, we would say a compressed file stored in a computer is a (partial) copy of the original work.[102] But the file, while an imperfect copy, is still deterministic, not probabilistic: the same file will generate the same imperfect copy every time.[103] A given model stored on a computer is also fixed; its weights don't change from one generation to the next. But because those weights encode probabilistic relationships, even where a model has memorized training data, extraction may in some cases occur only probabilistically: the same prompt may sometimes elicit the memorized material and sometimes not.

---

97. The definition of a computer program might also seem relevant here. A computer program is "a set of statements or instructions to be used directly or indirectly in a computer in order to bring about a certain result." *Id.* While that would seem to encompass model weights that generate a copyrighted output, the fact that the definition is nowhere connected to either copyrightability or infringement makes it less useful as a guide. *See generally* Cooper & Grimmelmann, *supra* note 9.

98. § 101.

99. *See* NAT'L COMM'N ON NEW TECH. USES OF COPYRIGHTED WORKS (CONTU), FINAL REPORT OF THE NATIONAL COMMISSION ON NEW TECHNOLOGICAL USES OF COPYRIGHTED WORKS 44 (1978) [hereinafter CONTU REPORT] ("In every case, the work produced will result from the contents of the data base, the instructions indirectly provided in the program, and the direct discretionary intervention of a human involved in the process."). CONTU rejected the idea that the computer itself could generate new content as "too speculative to require serious consideration." Pamela Samuelson, *Allocating Ownership Rights in Computer-Generated Works*, 47 U. PITT. L. REV. 1185, 1195 (1986) (citing CONTU REPORT, *supra*, at 44). Samuelson notes that even in the 1980s that was incorrect in some cases: "Although both the programmer and user might contribute to the framework within which the computer makes its selections or arrangements of data, the computer actually makes the selection." *See id.* at 1199, 1201.

100. Samuelson, *supra* note 99, at 1202 ("[T]he user will have been the instrument of fixation for the work, that is, the person who most immediately caused the work to be brought into being.").

101. For a discussion of lossy compression in the AI context, see Cooper & Grimmelmann, *supra* note 9, at 188–89; and *supra* Section II.B.

U.S. copyright law hasn't really had to confront whether copyrighted material that can be elicited from a computer system only probabilistically—appearing in some outputs but not others depending on the probabilities of each successive word appearing—is "fixed" in the sense required to qualify as a copy.[104] A computer program compiled from source code into object code (the

102. Many familiar file formats, like JPEG for images, are lossy. As a matter of information theory, it is an interesting problem how much information must be provided to "contain" a work. But this is out of our present scope. For a discussion, see generally Sarah Scheffler, Eran Tromer & Mayank Varia, *Formalizing Human Ingenuity: A Quantitative Framework for Copyright Law's Substantial Similarity*, *in* CSLAW '22: PROCEEDINGS OF THE 2022 SYMPOSIUM ON COMPUTER SCIENCE AND LAW 37 (Ass'n for Computing Mach. 2022).

103. Language modeling is compression in a precise information-theoretic sense: a probabilistic language model, paired with an appropriate coding scheme, induces a lossless compressor. This does *not* mean the model is a lossless archive of its training data. Rather, the claim is *distributional*: data drawn from the kind of distribution the model captures can be efficiently encoded *on average*. The underlying intuition is simple: predictable material takes less information to describe, and is therefore more compressible. So the better the model is at predicting what comes next, the fewer bits are needed to encode the resulting text, on average. In this sense, an LLM is an effective compressor because of its predictive capabilities. Importantly, this is *not* a claim that every piece of training data is encoded in a lossless or retrievable form, such that it can be reconstituted verbatim or near-verbatim in generation. Rather, verbatim and near-verbatim extraction reflects a specific, high-fidelity form of compression that is a hallmark of memorization, not of language modeling more generally. *See supra* Section II.B; Chiang, *supra* note 54, at 5; Getty Images (US) Inc v. Stability AI Limited [2025] EWHC (Ch) 2863 [555] (Eng.) ("[T]he model weights do not directly store the pixel values associated with billions of training images . . . . A single training image has only a limited impact on how the parameters of the network will be changed." (internal quotation marks omitted)). *See generally* Delétang et al., *supra* note 29.

104. A Munich court decision found that memorization did create a copy under EU copyright law, see Landgericht München I [LG München I] [Munich Regional Court I], Nov. 11, 2025, Gewerblicher Rechtsschutz und Urheberrecht [GRUR] 1917 (2025) (Ger.), though that decision is currently under appeal. *See, e.g.*, Zbigniew Okoń, *Memoryzacja utworów przez modele AI w świetle unijnego prawa autorskiego* [*Memorization of Copyrighted Works by AI Models Under EU Law*], EUROPEJSKI PRZEGLĄD SĄDOWY [EUROPEAN JUD. REV.], no. 12, 2025, at 19 (Pol.), *translated in* Zbigniew Okoń, Memorization of Copyrighted Works by AI Models Under EU Law (Feb. 10, 2026) (unpublished manuscript), https://papers.ssrn.com/sol3/papers.cfm?abstract_id=6041395 (on file with 1s and 0s that actually operate the program) doesn't look anything like the source code that generated it. But the relationship between the two is designed to be deterministic: the object code is the predictable result of compiling that source code with a particular compiler. That determinacy is what makes the compiled program reliably produce the behavior one would expect from the written source code. Even versions that require more judgment, like language translations, are nonetheless instantiated in a permanent form: once made, the relationship between the original and the translation doesn't vary from one access to the next.

The closest cases to present issues of non-determinism involve gardens and video games. In *Kelley v. Chicago Park District*,[105] the Seventh Circuit held that a garden was not copyrightable because it wasn't fixed, though a blueprint design for a garden could be.[106] The court found that "a garden is simply too changeable to satisfy the primary purpose of fixation; its appearance is too inherently variable to supply a baseline for determining questions of copyright creation and infringement."[107] The court reasoned that because gardens grow and change organically, their evolution is unpredictable.[108]

We think *Kelley* goes too far. While gardens certainly do grow and change in ways that are not fully predictable, both the initial conditions and the way the gardens are curated will significantly affect those changes in a way that seems likely to be mostly if not entirely reproducible. Certainly, the fact that people plan gardens with blueprints suggests that gardens have a design that, if not fully deterministic, is implementable and sufficiently replicable to be usable.

In some prior cases, we see a limited form of non-deterministic behavior in the output of a computer program. In video games, for example, the game code generates images that interact with human behavior to produce gameplay sequences on the screen that are not stored in the computer. My experience of a combat sequence, for example, will differ from yours, because we won't hit the same keys in the same order. Nonetheless, we treat those audiovisual sequences as copies of the underlying work despite the interactivity of the game, which

---

the Berkeley Technology Law Journal). *See generally* Matthias Leistner & Lucie Antoine, *TDM and AI Training in the European Union—From 'LAION' to Possible Ways Ahead?*, 74 GRUR INT'L 1027 (2025) (Eng.). By contrast, an English court, evaluating a different model and modality (images) with different evidence presented, has held the opposite. *See Getty Images*, [2025] EWHC (Ch) 2863 [603].

105. 635 F.3d 290, 304 (7th Cir. 2011).

106. *Id.* at 304–05. *Kelley* treats fixation as a matter of copyrightability, not infringement. But since the standards are the same and rely on the same statutory definition, presumably it would apply to the fixation required for infringement, as well.

107. *Id.*

108. *Id.* at 305 ("Although the planting material is tangible and can be perceived for more than a transitory duration, it is not stable or permanent enough to be called 'fixed.' Seeds and plants in a garden are naturally in a state of perpetual change; they germinate, grow, bloom, become dormant, and eventually die. This life cycle moves gradually, over days, weeks, and season to season, but the real barrier to copyright here is not *temporal* but *essential.* The essence of a garden is its vitality, not its fixedness. It may endure from season to season, but its nature is one of dynamic change.").

means that the precise sequence of images that appear on a screen may not have been generated or even contemplated when the game was written.[109]

In *Micro Star v. Formgen Inc.*, for example, the court held that a user of a video game infringed by profiting from new map files that called on the game assets to generate new levels of the game.[110] The maps were new, so the combination of images the user created didn't exist in the original game. Nonetheless, the court held that the map files infringed because they provided instructions to a computer that created new gameplay using copyrighted images from the original game.[111] That is a form of non-deterministic output that is

109. Although there is player interaction with the machine during the play mode that causes the audiovisual presentation to change in some respects from one game to the next in response to the player's varying participation, there is always a repetitive sequence of a substantial portion of the sights and sounds of the game, and many aspects of the display remain constant from game to game regardless of how the player operates the controls. *See* Stern Elecs., Inc. v. Kaufman, 669 F.2d 852 (2d Cir. 1982) (finding fixation on the screen display of a computer game because even though the images displayed were not themselves stored anywhere in the computer, "many aspects of the sights and the sequence of their appearance remain constant during each play of the game," and "[t]he repetitive sequence of a substantial portion of the sights and sounds of the game qualifies for copyright protection as an audiovisual work"); Williams Elecs., Inc. v. Artic Int'l, Inc., 685 F.2d 870, 874 (3d Cir. 1982); *accord.* Midway Mfg. Co. v. Artic Int'l, Inc., 704 F.2d 1009, 1012 (7th Cir. 1982) ("The player of a video game does not have control over the sequence of images that appears on the video game screen . . . . The most he can do is choose one of the limited number of sequences the game allows him to choose.").

110. *See* 154 F.3d 1107, 1114 (9th Cir. 1998).

111. *Id.* at 1112. The court held that the maps were a derivative work of the underlying game. This was a bit unusual, because the maps didn't incorporate any features of the game. But they were encoded to generate gameplay that called on and infringed the copyright in the original game.

A number of copyright plaintiffs in generative AI cases have alleged that even if the model isn't a copy of their works, it is a derivative work. Those arguments have largely been rejected by the courts. *See, e.g.*, Andersen v. Stability AI Ltd., 700 F. Supp. 3d 853, 865–66 (N.D. Cal. 2023) (dismissing plaintiff's claim that the model itself is an infringing derivative work); Kadrey v. Meta Platforms, Inc., No. 23-cv-03417-VC, 2023 WL 8039640, at *1 (N.D. Cal. Nov. 20, 2026) (granting motion to dismiss, describing that "[t]he plaintiffs allege that the LLaMA language models are themselves infringing derivative works because the models cannot function without the expressive information extracted from the plaintiffs' books" and that "[t]his is nonsensical" (internal quotation marks omitted)); *see also* Tremblay v. OpenAI, Inc., 716 F. Supp. 3d 772, 778 (N.D. Cal. 2024) (dismissing the plaintiffs' allegation that "every output of the OpenAI Language Models is an infringing derivative work" (internal quotation marks omitted)); Justice v. Uncharted Labs, Inc., No. 25 Civ. 5026 (AKH), 2026 WL 1430232, at *2 (S.D.N.Y. May 21, 2026) (dismissing plaintiffs' allegations that the outputs of a model were derivative works because the model was originally trained on those works). That's because the law has been interpreted to say that a work is a derivative work only if it is substantially similar in protectable expression to the original work—the same standard for making a copy. *See* Litchfield v. Spielberg, 736 F.2d 1352, 1357 (9th. Cir. 1984) (stating that a work is a derivative work only if "it would be considered an infringing work if the material which it has derived from a prior work had been taken without the consent of a copyright proprietor of such prior work," and "[t]o prove infringement, one must show substantial similarity" to that prior work (emphasis removed) (quoting United States v. Taxe, 540 F.2d 961, 965 n.2 (9th Cir. 1976))). That means plaintiffs can't use the derivative works right as an end-run around the limits of the right to make copies.

Some have called the derivative works right "completely superfluous" as a result. *See* 2 MELVILLE B. NIMMER & DAVID NIMMER, NIMMER ON COPYRIGHT § 8.09[A][1] (2026). But that's not entirely correct. A derivative work can involve a change to a physical copy of a copyrighted work rather than somewhat closer to what we see with generative AI. But it is nonetheless different in two important respects.

First, the content produced by playing a video game may not be contained in the same form in the game itself, but the content it will output is generally deterministic. I can make my character look different than others, but I can only do it by combining exact tools and images in the game itself. And I can make that character do many different things in different places in the game, but each of those actions and places are dictated by the parameters of the game. I can't take my character to a place that doesn't exist in the game or have them do something the rules of the game don't permit.

Games like *No Man's Sky* with procedurally generated levels present a harder question for fixation. In procedurally-generated games, the levels are generated anew—involving a degree of randomness—as the player arrives at a new planet. They don't exist until the player arrives, so they aren't stored in the computer, but they're generated using a known algorithm working from a fixed base of content. There isn't litigation on this, but it seems likely a court would say the procedurally generated levels (much less all the possible ones that *might* be generated) didn't preexist in the game; they were created when the user arrived at a new planet and the game engine generated a planet for them. Thus, planets that might at some later point be created in a game aren't fixed in the game itself.

The non-deterministic fixation issue is unlikely to arise in practice in most video game cases because of the second important distinction: the content I generate when I play a video game is the *output* of the interaction between me and the game, displayed or recorded in some medium. It isn't in the game itself. The reason it infringes is that the output, while new and even unpredictable in a sense, copies identifiable elements of the game that *are* present in the computer in a fixed, deterministic form. That was true in *Micro Star*, and it would be true of *No Man's Sky*. We don't need to ask whether the new levels are somehow "in" the creation of a new work. *See, e.g.*, 17 U.S.C. § 101; Mirage Editions, Inc. v. Albuquerque A.R.T. Co., 856 F.2d 1341, 1342–43 (9th Cir. 1988).

Another possible difference—one that is relevant for our purposes—is that the statute nowhere says that a derivative work must be fixed, as a copy must. *See* § 101 (defining derivative work as "a work based upon one or more preexisting works, such as a translation, musical arrangement, dramatization, fictionalization, motion picture version, sound recording, art reproduction, abridgment, condensation, or any other form in which a work may be recast, transformed, or adapted"). Nonetheless, courts have held that there must be some sort of quasi-fixation to create a derivative work. *Micro Star*, 154 F.3d at 1111 (reaffirming that because "all the Copyright Act's examples of derivative works took some definite, physical form and . . . this was a requirement of the Act"); Lewis Galoob Toys, Inc. v. Nintendo of Am., Inc., 964 F.2d 965, 967 (9th Cir. 1992) ("A derivative work must incorporate a protected work in some concrete or permanent 'form'"). Were it otherwise, acts like speaking aloud or reading an email on a computer screen could be viewed as the creation of ephemeral derivative works. *See* Ochoa, *supra* note 89, at 262 (reading the statute to not require fixation "would render illegal merely imagining a modified version of a copyrighted work"); *cf. Talkin'*, *supra* note 20, at 327–28 (noting but not resolving this issue).

the game in fixed form, because it's the act of generating them as output that is the infringing use.

Contrast a video game with a word processing program. Microsoft Word contains all the elements necessary to build any text output, including this Article, all seven *Harry Potter* books, and an infinite number of works that have never been and may never be created. But we wouldn't say that those works are "in" Word merely because they can be assembled from components present in Word. Those individual components (letters, numbers, punctuation, and even higher-order operations like spelling and grammar correction) aren't copyrightable. It's the act of combining them into a creative output that generates a copyrightable work.

But even if that weren't true, it would be nonsensical to say that Word somehow contains all works ever written or that could be written.[112] It may contain the tools for creating those works, but that doesn't mean it has a copy of the works themselves, any more than the digits of pi or any other infinite sequence necessarily contain *Hamlet* (and all other works of authorship now known or later conceived). A key difference here is what information we need as an input in order to retrieve a particular output. In a deterministic file storage system, the address or perhaps the name of the file is enough to retrieve the content of that file. At the other extreme, to say that Microsoft Word or the alphabet encodes *Hamlet* or *Harry Potter* isn't meaningful because they aren't there in any ordered form from which they reasonably can be retrieved. Yes, the works are "in there" in some, very generous, sense, but you would need to actually write the works in order to retrieve them; you'd have to input the information in order to output it.

### *C. Applying Non-Deterministic Analysis to Generative AI*

Generative AI sits in an odd space between these examples. LLMs generate text on the basis of probabilistic word and context relationships that are encoded in their model weights.[113] Those relationships are influenced by the works the models train on. So an LLM encodes certain word relationships as grammatical and others as nonsensical based on its training. But an LLM isn't simply a neutral collector of word relationships. The more heavily particular text strings are

112. *Cf.* Samuelson, *supra* note 99, at 1208 & n.94 ("Granting all rights to the programmer would mean that the programmer would automatically own everything the program was *capable* of generating . . . . Because it is possible to put the entire English language into a computer program that can construct prose, it would theoretically be possible for the programmer to claim rights in all writings that were—or could be—produced by the program. Such a claim may seem ridiculous, and yet it would be a logical result of automatically awarding rights in whatever the program did or could produce." (footnote omitted) (citation omitted)). While Samuelson's focus was on the establishment of rights, fixation is symmetrical, and the same logic applies to the making of fixed infringing copies.

113. Sag, *supra* note 39, at 24 ("The knowledge that resides within an LLM is inherently probabilistic."); *see* Cooper & Grimmelmann, *supra* note 9, at 197 ("[T]he probabilistic element . . . is inescapable.").

represented in training data, the more likely the LLM will be to replicate those word strings verbatim or nearly verbatim.[114] Consider Figure 1: in general, there are many plausible, grammatical English-language continuations of "Mr. and Mrs. Dursley"; after all, these are just two names. The fact that Llama 3.1 70B assigns overwhelmingly high probability to the verbatim continuation from the opening line of *Harry Potter and the Sorcerer's Stone* is not explained by any feature of the English language, but by the model's memorization of that specific passage from its training data.

We have shown that for certain works, like *Harry Potter and the Sorcerer's Stone*, *1984*, and many others, it's straightforward to get some LLMs, notably Meta's Llama 3.1 70B, to deterministically generate all or a substantial part of the work essentially verbatim with minimal prompting.[115] These works are highly memorized by Llama 3.1 70B; the extraction probabilities across each text are over 95% in all locations,[116] which we found makes deterministic extraction straightforward in practice. That suggests that, even if there isn't a copy of those works stored in Llama 3.1 70B in the same sense that we would store a Word document in a computer file, the model has stored a set of instructions from which it's straightforward and deterministic to generate the work as output. And that result is repeatable; running the same queries produces the same basic result. If that isn't exactly like a computer program stored in memory, it's close enough that a court would probably say that there's a copy (or at least near-perfect blueprints for the making of a copy) in the model itself.[117]

114. *See Deduplicating*, *supra* note 11, at 8430 ("Data duplication has the effect of biasing the trained LM towards particular types of examples. This can contribute to a lower diversity of generations, and increased likelihood that the generated content is copied from the training data.").

115. *See supra* Section II.A (discussing decoding schemes); *supra* Section II.B (discussing greedy vs. deterministic extraction). We used a deterministic decoding algorithm called beam search.

116. *See supra* note 68 and accompanying text. This 95% number applies to contiguous sequences of 50 tokens: there for a given prompt of 50 tokens, there is a 95% probability of outputting the verbatim following 50 tokens. With respect to 50-token inputs and outputs, such high extraction probabilities exist throughout these works. Notably, we did not observe the same amount of memorization of these books even in other Meta models, including Llama 1 65B and Llama 2 70B.

117. There are many other ways to extract training data that fit this general description, but they all ultimately involve some form of prompting. Ahmed et al. extracted large portions of near-verbatim text from production LLMs—models that are aligned and embedded inside systems, that are interacted with through an API. Extracting Books (Production LLMs), *supra* note 13, at 12, 14, 18–19. In some cases, they had to "jailbreak" the models with adversarial prompts to extract training data (e.g., for Claude 3.7 Sonnet and ChatGPT 4.1); in others, no jailbreaking was necessary (Gemini 2.5 Pro and Grok 3). *Id.* at 4. Fine-tuning a model on additional data can also expose memorization of data it was previously trained on, for reasons that aren't particularly well understood. *See generally* Nasr et al., *supra* note 12 (finding that fine-tuning GPT 4 makes it possible to extract additional pre-training data). More recently, others have shown that fine-tuning on books can enable extraction of other copyrighted books the model was previously trained on, demonstrating memorization of those books in the model. *See generally* Xinyue Liu, Niloofar Mireshghallah, Jane C. Ginsburg & Tuhin Chakrabarty, Alignment Whack-a-Mole: Finetuning Activates Verbatim Recall of Copyrighted Books in Large Language Models (Mar. 28, 2026) (unpublished manuscript), https://arxiv.org/pdf/2603.20957 [https://perma.cc/TF75-PHE2]. This type of fine-tuning-based result requires changing the model through additional training, such that the weights are no longer the same as those that are available in systems deployed to This is also the case for many other works that aren't memorized to quite the same extent—for example, those that take multiple tries to extract or for which we can extract only isolated fragments of the work. In many cases, deterministic decoding methods can extract works near-verbatim where the extraction probability is only 0.1%.[118] The point is, a particular work may be more weakly memorized than another in a model, but it may still be possible to retrieve both works reliably and with just one prompt to the model.[119]

For other works, the opposite seems to be true. Many models won't output all or any significant part of many works on which we know they trained, even if we try thousands or tens of thousands of times, as with our experiments on *Sandman Slim* (Figure 3). It seems clear that there isn't a copy of those works in the model in any meaningful legal sense, at least based on what we can determine using current extraction methods. The model may have been trained on those works, but it apparently didn't retain information about them sufficient to be able to extract them in whole or in part.[120]

There's also a middle class of works for which it may only be possible to extract significant fractions of the work probabilistically, rather than deterministically.[121] We may need to run hundreds or thousands of prompts before the work is extracted.[122] In this case, we would be seeing non-determinism in action. The model would generate tokens based on its encoded probability of those tokens following existing ones, based on what it has learned from training. But it wouldn't generate the same result every time. So the model is like an enormous decision tree, where each token position is a new branching point that can select a different possible token. The model may follow different branches of the tree each time; only the particular branch that happens to the general public. Some systems, like ChatGPT, expose fine-tuning APIs that enable such customization.

118. *See generally* First Principles, *supra* note 65.

119. In the distribution the model assigns, 0.1% may be close to the highest probability sequence for a given prompt; deterministic decoding algorithms like beam search, which approximate identifying the highest probability sequence, will therefore often return such sequences. *Near-Verbatim*, *supra* note 13, at 4–5; *see generally* First Principles, *supra* note 65.

120. We say "apparently" because, as we noted above, our methods for extracting text and therefore proving memorization in models are imperfect and in development. That is especially true for closed-source models for which scholars need to use adversarial extraction techniques. *See, e.g.*, Extracting Books (Production LLMs), *supra* note 13, at 4. Indeed, our own prior work has significantly expanded the universe of provable memorization. *See generally Extracting Books (Open-Weight)*, *supra* note 12; *Near-Verbatim*, *supra* note 13. It's possible, and even likely, that methods developed later will be able to extract more training data than we can with today's methods.

121. These would be cases where using a deterministic algorithm like beam search would fail to return the memorized sequence. *See supra* note 115 and accompanying text. In this case, we could use a stochastic decoding algorithm that introduces non-determinism in outputs.

122. These training data are still memorized in a technical sense, but, intuitively, are not memorized as strongly as in the cases where we can extract the work in just a handful of attempts.

correspond to the book will generate text from the book.[123] The other branches may generate content that is substantially similar to the book, or they may generate new and previously unknown content.[124] Many of the outputs generated may not be infringing.[125]

In this particular set of circumstances, it may be hard to say there is a "copy" of the work in the model in the sense the Copyright Act means the term without saying there is a "copy" of every branch of the enormous decision tree, most of which don't actually exist in the outside world. From one perspective, if it takes 1,000 tries at a prompt to generate verbatim output, the 999 tries that are *not* copies aren't somehow pre-stored in the AI model.[126] To say otherwise, we would have to say that a model small enough to run in a single data center (or sometimes on a single computer) somehow stores more unique outputs than there are atoms in the observable universe. That's akin to saying that Microsoft Word stores all the things that could be written with it in a "tree" of potential letters following each other. Yes, it's possible to generate those things from the model, but it doesn't mean they are all sitting in the model already made. They are being generated on the fly. But, if that's true of noninfringing outputs generated by the probability tree, it seems wrong to say the result is any different for the one output that happens to infringe. The output, if and when it's generated, may be a copy, but the untapped potential for generating it isn't a copy.

But that conclusion may depend on the underlying sequence probabilities. One possibility is that the memorized output has an extraction probability of about 1/1,000 (0.1%), and the other 999 non-memorized outputs might have substantially lower generation probabilities (e.g., 1/100,000,000, or 0.000001%). If we took another sample of 1,000 tries, we would perhaps return the memorized sequence again once and 999 very different non-memorized outputs of similarly low probability. There's clearly something meaningfully different between the single probabilistic extraction case and the other outputs the model generates,

123. This picture is also consistent with *supra* Figure 1. It's just that one branch (the memorized one) is so high probability that common decoding algorithms mean it's possible to deterministically generate the text.

124. *See Near-Verbatim*, *supra* note 13, at 9–10.

125. As our work has shown for the verbatim case, this is an open empirical question that depends on the model and work. *Extracting Books (Open-Weight)*, *supra* note 12, at 18. In some cases, most of the outputs may be noninfringing. In others, a branch of (near-)verbatim extraction with 1/1,000 probability may be deterministically decodable, depending on how low the probabilities are of the other branches. (If 1/1,000 is by far the highest probability branch in the tree, common deterministic decoding algorithms, like beam search, will return it every time.) So even the 1/1,000 chance number carries important nuance. See *supra* Section II.A for more on decoding; and Cooper & Grimmelmann, *supra* note 9, at 148–56.

126. This assumes the other tries aren't also similar enough to be infringing; most of the near-verbatim text might be treated the same as the verbatim text for these purposes. So our work showing that there is more near-verbatim memorization than verbatim memorization expands the universe of cases that look more like *Harry Potter*. *See Near-Verbatim*, *supra* note 13, at 1. For additional nuances for this 1/1,000 number, see *supra* note 119; *see also* Cooper & Grimmelmann, *supra* note 9, at 194–97.

even if it takes on average 1,000 tries to extract the memorized training data. From this perspective, like the other, it's also clear that the 999 tries that are *not* copies aren't somehow pre-stored in the AI model; but it's less clear that the one try that yields the memorized work isn't a copy.

If this type of probabilistic case were deemed not a copy, that doesn't mean the model hasn't memorized the work. The model has been trained on a work, has clearly learned things from that work, and it's possible that it will produce that work—much more possible than random chance.[127] From a machine learning perspective, it may make sense to say the work is "memorized."[128] But that doesn't mean the model contains a deterministic copy of that work that is fixed in the model "for a period of more than transitory duration," as the statute requires.[129] The information is there. It's possible with some significant effort to extract it in the order required to replicate a copyrighted work. But if that extraction is just one possibility among thousands, the work is produced at the moment of extraction. The information necessary to create that extraction is present in the system. But, for probabilistic cases where it may take thousands of attempts to realize the memorized training data in outputs, there is likely no copy as the law uses the term.

Generation on the fly in response to prompts likely isn't fixation for a period of more than transitory duration. The legislative history excluded from this concept of fixation embodiments that were "captured momentarily in the 'memory' of a computer."[130] The Ninth Circuit ignored this part of the statute in

127. An old adage holds that if you give enough monkeys enough typewriters, they will eventually produce the works of Shakespeare by chance. So some presume that, similarly, given enough attempts a model might generate by happenstance text identical to a copyrighted work even if it hadn't memorized that work. But LLMs are not like the monkey at the typewriter; the monkey hits keys randomly, but the LLM has learned a probability distribution that isn't completely (uniform) random over text. Further, our work has shown that the kinds of probabilistic copying we discuss are many, many orders of magnitude more likely than could be explained by the model's general ability to produce text. *See Extracting Books (Open-Weight)*, *supra* note 12, at 10–11; First Principles, *supra* note 65, at 1. To do so, we can measure the probability of a model generating *non*-training data (i.e., a case for which memorization is impossible). By doing this for books that are closely matched to known training data (e.g., sequels in the same series that were published after the model was trained), we have found that non-training-data generation probabilities range from $10^{-27}$ to $10^{-16}$ (where, as a point of reference, $10^{-3}$ is 0.1%). As Cooper et al. note: "To give a sense of just how small [the larger of the two numbers ($10^{-16}$)] is, the chance of winning the Powerball lottery jackpot in a single draw is about 1 in 292 million—roughly 34 million times more likely." *Extracting Books (Open-Weight)*, *supra* note 12, at 10.

128. *See* Cooper & Grimmelmann, *supra* note 9, at 177–78; *see also Extracting Books (Open-Weight)*, *supra* note 12, app. E (providing a robust appendix). Cooper and Grimmelmann say that "the probability of generating a near-exact copy of a piece of training data has to be high enough to rule out coincidence." Cooper & Grimmelmann, *supra* note 9, at 198. That is right if the goal is to show memorization. And as they note, a 1/1,000 chance of generating the work in response to a prompt that tries to elicit it is far, far more unlikely than a truly random chance. *Id.* at 198 n.212. But the fact that information latent in the system is sufficient to generate part of all of the work is different than saying that a copy persists in the system.

129. 17 U.S.C. § 101.

130. H.R. REP. NO. 94–1476, at 53 (1976). Similarly, projecting an image on a movie screen for 1/24th of a second is enough for us to perceive it, but clearly doesn't count as fixation. *See id.*

*MAI Systems Corp. v. Peak Computer, Inc.*,[131] but both the Second and Fourth Circuits have now held that buffering for a few seconds does *not* create a copy that exists for more than a transitory duration, even though it is long enough to be perceived.[132] Permanent storage in computer memory is fixation. And outputting a copy of the work creates a new, fixed copy. But the process of generating results on the fly in the computer doesn't fix a copy in the model itself.

Our ultimate (and tentative) conclusion is thus mixed. We view a sufficient ease of generation as evidence that the work as a whole is in fact latent in the model and not just being generated by chance. The ability to deterministically generate the first *Harry Potter* book using Llama 3.1 70B from a simple seed prompt of ground-truth text from the book suggests that the model has stored instructions that correspond sufficiently to the book, such that it's akin to the storage of a computer program or video game content. But for books that can be extracted only through a high number of efforts producing different outcomes, what is stored is not a copy of the book itself, but a set of ingredients that can produce copyrighted content only when combined in just the right way. A collection of ingredients is not an end product.

That conclusion means that we cannot simply say categorically that models do or do not contain copies of a work in the copyright sense. Some models contain copies of some works under our reading of the statute.[133] No models (so far as we know) contain copies of all the works on which they are trained. This is possible in degenerate cases that don't reflect state-of-the-art training practices;[134] but it's impossible for models like Llama 3.1 70B, which are trained on datasets that are far too large for every piece of training data to be losslessly compressed in a model of that size. Ultimately, the question is an evidentiary one that may differ for each model and for each work in that model's training data.[135] Furthermore, the answer may change, both as new models are developed (ours and others' research suggests, not surprisingly, that larger models memorize

131. *See* 991 F.2d 511, 518 (9th Cir. 1993).

132. Cartoon Network LP v. CSC Holdings, Inc., 536 F.3d 121, 128–30 (2d Cir. 2008); CoStar Grp., Inc. v. LoopNet, Inc., 373 F.3d 544, 556 (4th Cir. 2004).

133. Carlini et al. estimated that a 6 billion parameter model memorizes 1% of its dataset. Carlini et al., *supra* note 78, at 1. Other estimates range from 0.1% to 10% depending on the model and the technique used. *See Deduplicating*, *supra* note 11, at 8430; Nasr et al., *supra* note 12, at 15. Larger models may memorize more, and previous tools for identifying memorization may undercount it, so that number is likely to increase.

134. *See* Cooper & Grimmelmann, *supra* note 9, at 199.

135. *Id.* at 190–91; *see also Carlini Blog*, *supra* note 15, at 3 ("[M]odels sometimes emit training data. It doesn't happen every time. Models can produce output that is not just a direct copy of what they were trained on. And it doesn't happen never. But sometimes, models directly output text or images that they were trained on.").

more than smaller ones)[136] and as new ways of assessing the ease of extraction are developed, producing new evidence.

### *D. The Consequences of Defining a Copy*

The consequences of concluding that a model contains a copy of a work as the Copyright Act defines that term are potentially significant.

First, if we conclude the model contains a fixed copy and doesn't just generate one when prompted, that copy is prima facie infringement whether or not it's ever output or used in any way.[137] Whatever copy exists in the model can be "reproduced" even if it's never "perceived" or "communicated."[138] As a policy matter we might not want to include copies that can never be viewed or used as within the scope of copyright,[139] but courts have generally rejected the "communicative function" theory of fixation.[140] So, if we decide that anything that might be memorized is a copy for legal purposes, the model itself is an unlawful copy of each of those works unless protected by a defense or exception.

Further, a copy doesn't have to be of the entire work to infringe. Copying even a small part of the work is sufficient as long as the copy includes a more than de minimis amount of protectable expression.[141] So while our prior work has shown that for many works it's possible to extract only some specific portions, that fact doesn't mean a model wouldn't infringe that work. If we conclude that the possibility of partial extraction makes the model a copy of those parts of the work, the model would still infringe copyright by copying those

136. *See Extracting Books (Open-Weight)*, *supra* note 12, at 41; Carlini et al., *supra* note 78, at 2633; Hayes et al., *supra* note 68, at 9273; *Near-Verbatim*, *supra* note 13, at 82.

137. *See* Walker v. Univ. Books, Inc., 602 F.2d 859, 864 (9th Cir. 1979) ("[T]he fact that an allegedly infringing copy of a protected work may itself be only an inchoate representation of some final product to be marketed commercially does not in itself negate the possibility of infringement."); Sega Enters. Ltd. v. Accolade, Inc., 977 F.2d 1510, 1518 (9th Cir. 1992) ("[T]he Copyright Act does not distinguish between unauthorized copies of a copyrighted work on the basis of what stage of the alleged infringer's work the unauthorized copies represent."); MAI Sys. Corp. v. Peak Comput., Inc., 991 F.2d 511, 518–19 (9th Cir. 1993); Thoroughbred Software Int'l, Inc. v. Dice Corp., 488 F.3d 352, 359–60 (6th Cir. 2007) (calculating damages for unused copies of software).

138. 17 U.S.C. § 101.

139. *See, e.g.*, Jessica Litman, *Fetishizing Copies*, *in* COPYRIGHT LAW IN AN AGE OF LIMITATIONS AND EXCEPTIONS 107 (Ruth L. Okediji ed., 2017); UMG Recordings, Inc. v. Shelter Cap. Partners LLC, 718 F.3d 1006, 1019 (9th Cir. 2013) ("[T]he language of the statute recognizes that one is unlikely to infringe a copyright by merely storing material that no one could access . . . [and suggests] that if the material were still being stored by the service provider, but was inaccessible, it might well not be infringing.").

140. Jay Dratler analogizes the "communicative function" argument to Count Dracula, because it "keeps emerging from the coffin, despite repeated interment." JAY DRATLER, JR., INTELLECTUAL PROPERTY LAW § 5.03[1][b], at 5–63 (1991); *see also* David G. Luettgen, *Functional Usefulness vs. Communicative Usefulness: Thin Copyright Protection for the Nonliteral Elements of Computer Programs*, 4 TEX. INTELL. PROP. L.J. 233, 237–38 (1996).

141. *See infra* Section III.D.

parts.[142] And even portions of works that can't be extracted verbatim can often be extracted with very small changes (i.e., near-verbatim).[143]

If models contain copies of some of the works they were trained on, as the law defines that term, the models themselves *are* infringing copies. That would mean that every new copy of a model also makes a copy of the copyrighted works contained in the model. That has significant consequences for open-weight models. For proprietary systems like those developed by OpenAI and Anthropic, those models exist behind closed doors. There are doubtless numerous copies of the model within the company, and each of those might be infringing, but the company doesn't widely distribute those actual models. By contrast, open-source business models that involve distributing copies of the model weights presumably are making new copies of the works whenever they make a copy of the model. So the potential damages exposure from copyright infringement may be greater for open-source models than for ones embedded in closed systems because of the way the models are used. That doesn't make much sense as a policy matter, since there is no reason to think open-source models are more problematic merely because of the way they are distributed, but it seems to follow from a strict application of the law.

Finally, the remedies for copyright infringement aren't limited to damages. If the model is a copy, it is presumably subject to remedies like injunctive relief and destruction.[144] There are good equitable reasons not to destroy the models, but plaintiffs have asked for it in numerous cases.[145] And while a less drastic remedy might seem to be removing the copyrighted works from the model, there's in fact nothing simple about "unlearning" in generative AI. As we have previously written, it's very difficult, perhaps even impossible, to entirely remove information from an existing copy of a model, and doing so can have unexpected negative consequences for other aspects of the model.[146] Nor is it

142. The exceptions will be cases where the things taken are not themselves copyrightable because they are too small, contain only quotes from other sources or uncopyrightable facts, or are otherwise unprotectable. For example, models that do not otherwise memorize portions of a book often memorize the copyright notice, because it appears repeatedly in standard form across multiple books. But memorizing that basic information would not be copyright infringement.

143. *Near-Verbatim*, *supra* note 13, at 1; Ippolito et al., *supra* note 52, at 29; *Extracting Books (Open-Weight)*, *supra* note 12, at 14–15, 184–85; Extracting Books (Production LLMs), *supra* note 13, at 32–33 (near-verbatim in production LLM outputs).

144. 17 U.S.C. §§ 502–503; *see also* Cooper & Grimelmann, *supra* note 9, at 169 (discussing models-as-copies and destruction).

145. *See, e.g.*, Complaint at 68, N.Y. Times Co. v. Microsoft Co., 777 F. Supp. 3d 283 (S.D.N.Y. 2025) (No. 23-cv-11195 (SHS)), Dkt. No. 1 ("Ordering destruction under 17 U.S.C. § 503(b) of all GPT or other LLM models and training sets that incorporate Times Works . . ."); Landgericht München I [LG München I] [Munich Regional Court I], Nov. 11, 2025, Gewerblicher Rechtsschutz und Urheberrecht [GRUR] 1917 (2025) (Ger.).

146. *See Machine Unlearning*, *supra* note 29, at 6–7.

evident that we can feasibly identify and come to voluntary licensing deals with those whose works turn out to be memorized in part.[147]

All of this assumes that the *prima facie* infringement that would result from declaring a model to contain a copy of a work leads to a finding of liability. But that conclusion doesn't necessarily follow, for two reasons.

First, it is possible that a copy in the model itself will be treated as *de minimis* and therefore noninfringing. To be infringing, the defendant must copy a more than de minimis amount of the work.[148] De minimis copying is normally thought of as copying a small amount of a work. That might apply in some cases of memorization, for example where a model memorizes only a short passage from a book.

But de minimis has another meaning in the law—not that the copying is of a small amount but that it is "trivial."[149] In deciding whether the copying is trivial, courts consider not just the amount copied but "the observability of the copied work—the length of time the copied work is observable in the allegedly infringing work," and factors such as "focus, lighting, camera angles, and prominence."[150] Indeed, courts have noted that "observability" is fundamental to the analysis.[151] When a work appears momentarily in the background of a film and is not the focus of the scene, for instance, so that the ordinary observer wouldn't notice it, courts may hold it de minimis even though the entire work is depicted.[152]

147. *See* Pamela Samuelson, Legally Speaking: Does Using In-Copyright Works as Training Data Infringe? 4 (July 15, 2025) (unpublished manuscript), https://papers.ssrn.com/sol3/papers.cfm?abstract_id=5342516 (on file with the Berkeley Technology Law Journal) ("Given the small and uncertain amount each book contributed to the models' value, transaction costs would make licensing infeasible."); Pamela Samuelson, *Justifications for Copyright Limitations and Exceptions* (discussing the various ways copyright markets fail to form or are dysfunctional, including transaction costs, disproportionate market power, regulatory interventions, and holdup problems), *in* COPYRIGHT LAW IN AN AGE OF LIMITATIONS AND EXCEPTIONS, *supra* note 139, at 12, 38; *cf.* Xiyin Tang, *The Class Action as Licensing and Reform Device*, 122 COLUM. L. REV. 1627, 1655 (2022) (noting the challenges of finding and licensing AI content).

148. Castle Rock Ent., Inc. v. Carol Publ'g Grp., Inc., 150 F.3d 132, 138 (2d Cir. 1998) (citing Ringgold v. Black Ent. Television, Inc., 126 F.3d 70, 75 (2d Cir. 1997)). *See generally* Oren Bracha, *Not De Minimis: (Improper) Appropriation in Copyright*, 68 AM. U. L. REV. 139 (2018) (collecting the law in every circuit).

149. Sandoval v. New Line Cinema Corp., 147 F.3d 215, 217 (2d Cir. 1998) (quoting *Ringgold*, 126 F.3d at 74); Gordon v. Nextel Commc'ns & Mullen Advert., Inc., 345 F.3d 922, 924 (6th Cir. 2003).

150. *Ringgold*, 126 F.3d at 75 (citing 4 MELVILLE B. NIMMER & DAVID NIMMER, NIMMER ON COPYRIGHT § 13.03[A][2] (1997)); *see also Gordon*, 345 F.3d at 924.

151. *Sandoval*, 147 F.3d at 217; Solid Oak Sketches, LLC v. 2K Games, Inc., 449 F. Supp. 3d 333, 344 (S.D.N.Y. 2020). One of us has shown that proprietary models vary widely in whether and how they build safeguards against extracting copyrighted works. Extracting Books (Production LLMs), *supra* note 13, at 19. For more on the systems vs. models distinction for extracting memorization, see Cooper & Grimmelmann, *supra* note 9, at 207–10.

152. *Ringgold*, 126 F.3d at 77; *cf.* Matthews Conveyor Co. v. Palmer-Bee Co., 135 F.2d 73, 85 (6th Cir. 1943) (asking "whether so much has been taken as would sensibly diminish the value of the original"); Gottlieb Dev. LLC v. Paramount Pictures Corp., 590 F. Supp. 2d 625, 632 (S.D.N.Y. 2008) (depicting a pinball machine in the background of a movie was de minimis).

Memorization in a model, even if it is determined to be a copy, may be de minimis if it's difficult to practically extract that copy from the model. Whether that's true is an empirical question that will differ from case to case. It might be difficult to extract a work as a practical matter because the probability of generating a work from a model is so low. If it takes 1,000 tries to extract a work, users probably won't persist merely to extract content they can already access elsewhere. Further, even if the work is heavily memorized in the model, that doesn't mean it would be easy to extract from a production system like ChatGPT that has some (however limited) guardrails built in to try to prevent surfacing copies in outputs to users. If the model actually deployed to consumers implements sufficient guardrails, it may be very difficult to extract the memorized work. And if you can't practically get the work out of the model, courts might view the inaccessible copy as de minimis. Courts haven't confronted such a case as a de minimis use, so it would be an extension of the doctrine, but it seems a logical one.[153]

Second, it's possible that the models, like the training datasets from which they were built, will be held a fair use of the works they memorize. U.S. courts have so far held that training models is fair use.[154] Even if a model contains a copy of some of the works it has trained on, that copy too might be fair use, especially if it's latent unless and until someone uses it to generate an infringing output. Many of the same arguments that justify fair use of the training data apply to memorization in the model itself.[155] If the work exists in the model but is never output by the model, so its presence there only affects the weights given to language the model uses when creating new and noninfringing works, the use of those copies is likely transformative for the same reasons it is in model training. And the market effects seem similar.[156] One difference is that the model itself is

153. *But cf.* Jessica Silbey & Samantha Zyontz, *De Minimis Copying: An Empirical Study*, 15 J. INTELL. PROP. & ENT. L. 118 (2025) (studying de minimis cases and finding that the argument usually fails).

154. Kadrey v. Meta Platforms, Inc., 788 F. Supp. 3d 1026, 1060 (N.D. Cal. 2025), Bartz v. Anthropic PBC, 787 F. Supp. 3d 1007, 1034 (N.D. Cal. 2025). *See generally* Mark A. Lemley & Bryan Casey, *Fair Learning*, 99 TEX. L. REV. 743 (2021); Sag, *supra* note 39. The Copyright Office took a different position, but it did so only by articulating a new theory of market harm that is directly at odds with existing caselaw. *Contrast* U.S. COPYRIGHT OFF., *supra* note 17, at 65–66 (reasoning that market dilution counsels against fair use), *with* Edward Lee, *Copyright Dilution Under Constitutional Scrutiny*, 25 CHI.-KENT J. INTELL. PROP. 1 (2026) [hereinafter Lee, *Copyright Dilution*] (explaining the flaws with the Copyright Office's argument). But only district courts have ruled on the issue, and the result could change. For a comprehensive analysis of training as fair use, see generally Edward Lee, *Fair Use and the Origin of AI Training*, 63 HOU. L. REV. 105 (2025).

155. *See* Tian "Tony" Chen, *Language Models' Verbatim Copying: Copyright Infringement Analysis Through the Lens of* The New York Times Co. v. Microsoft Corp., OpenAI, Inc. et al., 43 CARDOZO ARTS & ENT. L.J. 349, 364–66 (2025) (arguing that memorization in models should be fair use because it is non-expressive and serves the purpose of producing noninfringing outputs).

156. Some copyright owners and their advocates have suggested abandoning the existing legal definition of market effect and creating a theory of "market dilution," under which competing with a copyright owner would weigh the fourth factor against fair use, even if—as is normally the case—the competing work is not infringing. U.S. COPYRIGHT OFF., *supra* note 17, at 64–65; *see* Jacqueline C.

generally a commercial product (or the basis of a commercial service selling the use of the model), while the training dataset is not, and commerciality has taken on greater significance after *Andy Warhol Foundation for the Visual Arts, Inc. v. Goldsmith*, which revived it as an important consideration, rejecting thirty years of contrary precedent.[157]

Fair use by models has been much discussed elsewhere,[158] and it is not our intention to revisit that debate here. We raise it here only to note that the issue of whether a model contains a copy of a work isn't necessarily determinative. Where a model is deemed a copy, resolution of the fair use question takes on even more importance because of the legal consequences we discuss in this Section.

It's also possible that some (though not all) works that are now memorized might be licensed. But it's not feasible to simply create a license to all training content, for reasons Pam Samuelson has explained.[159] It also doesn't seem possible to eliminate all memorization,[160] and similar to "unlearning," even if it were possible, doing so might have negative unintended consequences for model behavior.[161] And it's not computationally feasible to determine which works each model memorizes in sufficiently relevant part in order to try to narrow down the universe of licenses.[162]

---

Charlesworth, *Generative AI's Illusory Case for Fair Use*, 26 VAND. J. ENT. & TECH. L. 323 (2025). That is not the law. Indeed, existing cases flatly reject it. *See, e.g.*, Sony Comput. Ent., Inc. v. Connectix Corp., 203 F.3d 596, 607 (9th Cir. 2000) ("[B]ecause the Virtual Game Station is transformative, and does not merely supplant the PlayStation console, the Virtual Game Station is a legitimate competitor in the market . . . . For this reason, some economic loss by Sony as a result of this competition does not compel a finding of no fair use."); Lee, *Copyright Dilution*, *supra* note 154.

157. *Compare* 143 S. Ct. 1258, 1285–86 (2023) (presuming commercial uses were unfair), *with* Campbell v. Acuff-Rose Music, Inc., 510 U.S. 569, 581 (rejecting the presumption that commercial uses were unfair). As Pam Samuelson notes, even after *Warhol* there are many ways commercial uses can be held fair. Pamela Samuelson, Lecture, *Justifications for Fair Uses*, 2025 WIS. L. REV. 1047, 1049–50 (2025).

158. *See generally* Peter Henderson, Xuechen Li, Dan Jurafsky, Tatsunori Hashimoto, Mark A. Lemley & Percy Liang, *Foundation Models and Fair Use*, 24 J. MACH. LEARNING RSCH., Sep. 23, 2023, art. 400; Cooper & Grimmelmann, *supra* note 9.

159. *See* Samuelson, *supra* note 147.

160. *See, e.g.*, Cooper & Grimmelmann, *supra* note 9, at 204, 204 nn.224–27; Vitaly Feldman, *Does Learning Require Memorization? A Short Tale About a Long Tail*, *in* STOC '20: PROCEEDINGS OF THE 52ND ANNUAL ACM SIGACT SYMPOSIUM ON THEORY OF COMPUTING 954, 958 (Konstantin Makarychev, Yury Makarychev, Madhur Tulsiani, Gautam Kamath & Julia Chuzhoy eds., 2020); *see also* Cooper & Grimmelmann, *supra* note 9, at 204–07; *Talkin'*, *supra* note 20, at 373 n.551; Sag, *supra* note 39, at 8 ("The answer is that avoiding memorization entirely is neither realistic nor desirable because there is no hard-edged distinction between memorization and learning.").

161. *See Machine Unlearning*, *supra* note 29, at 6–7.

162. It is feasible to test a potentially very large set of works, as we have done in some of our work. But training datasets are enormous, and it currently isn't feasible to test every model on every work in its respective training set. For instance, Cooper et al. spent over a year of GPU compute testing approximately 0.1% of the books in the Books3 corpus on Llama models, which are known to have been trained on Books3—and Books3 is only a small fraction of all of Llama's training data. *See generally Extracting Books (Open-Weight)*, *supra* note 12.

### *E. Policy Implications*

The Copyright Act was written in 1976. Its definition of "copy" was intended to encompass technologies that were just emerging at the time, including computers. But Congress didn't have in mind, and didn't address, the sorts of non-deterministic behavior we see with today's generative AI models and systems. In an ideal world, we might rewrite the law to limit copyright to covering copies that actually reach or communicate to people, deleting the word "reproduced" from the "perceived, reproduced, or otherwise communicated" language in the statute. That would exclude the actual models as copies, but would still encompass model output that is substantially similar to a copyrighted work.[163] But we don't live in an ideal world, and given the composition of Congress and the power of industry lobbyists, there are serious risks that revisiting the basic structure of the Copyright Act would end up making things worse, not better.

Given the indeterminacy of the Act's definition of "copy" and "fixed," perhaps courts could read the terms in ways that exclude the actual models as copies but would still encompass model output that's substantially similar to a copyrighted work. The closest analogy is to the cases that have treated temporary buffer copies internet sites make in the course of streaming or preparing to deliver content as not fixed and therefore not copies under the statute.[164] Those cases all involve temporary, intermediate instantiations of a work made in the service of delivery of that work. The instantiations in those cases mostly exist only for a short period of time, unlike the weights in AI models. But the courts seem to be applying the principle that a brief instantiation of a work in the service

163. Haviv et al. suggest that this is already the case, and that the law distinguishes memorization from copyright. Haviv et al., *supra* note 85, at 4–5. We are sympathetic to the policy argument, but skeptical that it reflects existing law, at least in the United States. Oren Bracha makes a similar argument. Oren Bracha, *The Work of Copyright in the Age of Machine Production*, 38 HARV. J.L. & TECH. 171, 179–80 (2024). He couches it as consistent with current law, but again, we are less convinced that is true. Courts have rejected the idea that a work must serve a communicative function to be copyrighted, applying the law to unpublished letters, for instance. *E.g.*, Salinger v. Random House, Inc., 811 F.2d 90, 94 (2d Cir. 1987), *superseded by statute*, Act of Oct. 24, 1992, Pub. L. No. 102-492, 106 Stat. 3145 (codified at 17 U.S.C. § 107); *cf.* Luettgen, *supra* note 140, at 250–52 (arguing for narrower protection for noncommunicative works but not contending that reflects current law). Under the statute the same should also be true of infringing copies, since the definition of the terms is identical for both sections. Nor is it the case, as Haviv et al. suggest, that memorization is generally of "generic, functional, or weakly protected material." Haviv et al., *supra* note 85, at 8. While that is surely sometimes true, we and others have shown numerous instances of verbatim memorization of creative works. See *supra* note 88 for additional discussion.

164. Cartoon Network LP v. CSC Holdings, Inc., 536 F.3d 121, 130 (2d Cir. 2008); UMG Recordings, Inc. v. Shelter Cap. Partners LLC, 718 F.3d 1006, 1016 (9th Cir. 2013) ("[T]he language of the statute recognizes that one is unlikely to infringe a copyright by merely storing material that no one could access . . . [and suggests] that if the material were still being stored by the service provider, but was inaccessible, it might well not be infringing."); CoStar Grp., Inc. v. LoopNet, Inc., 373 F.3d 544, 556 (4th Cir. 2004); *cf.* Case C-496/24, Stichting Onderhandelingen Thuiskopievergoeding v. HP Nederland BV, ECLI:EU:C:2026:296 (Apr. 16, 2026) (holding that streaming doesn't create a copy even though it requires temporary instantiations of the files).

of generating an end product that can't itself be perceived or used shouldn't be viewed as a fixed copy, and this seems analogous to generative AI. The parallel argument would be that even if research techniques reveal that a model contains a copy, if that copy is never perceived or communicated to the outside world, the law shouldn't condemn it. Alternatively, courts might treat such unobservable works as de minimis and therefore noninfringing.[165]

But either approach would require an extension of current law, either by statute or by judicial interpretation. As the law stands now, the answer to the question whether a model that memorizes content from a work is itself a copy of that work is: "it depends on the model and the work." That isn't a very satisfying answer, but it's the one the law and the science lead us to.

## IV.
## CONCLUSION

Memorization of copyrighted works in generative AI models forces copyright law to confront an issue it has so far managed to avoid—what to do about "probabilistic copies" of training works that the model might or might not generate in outputs. We think the most likely answer is a functional one: the model contains a copy of a work if, and only if, the work can be extracted from the model with relatively little effort. That means that whether a model contains a copy of a work will vary depending on the model, the work, and the state of the art of extraction research.[166] We would be better served by a rule that focused not on the metaphysical presence of copies in a model that may never be perceived in the outside world, but on whether the model generates infringing output.

165. It is also possible that a court will hold that the retention of some training data set in the model will not be viewed as a volitional act since no one decided to make that copy. On volition in AI, see Mala Chatterjee & Jeanne C. Fromer, *Minds, Machines, and the Law: The Case of Volition in Copyright Law*, 119 COLUM. L. REV. 1887, 1901–02 (2019); and Michael P. Goodyear, *Artificial Infringement*, UC L.J. (forthcoming 2026).

166. Indeed, over time, extraction science has developed new techniques that, in the instances in which they've been tested, make extraction easier—expanding our knowledge of how much state-of-the-art models memorize.